\documentclass[12pt]{article}

\usepackage{epsfig,latexsym,amsfonts,amsmath,amsthm,amssymb,amsbsy,multirow,slashed,wasysym,textcomp,dsfont,comment,mathtools,cancel,slashed,cite,datetime,appendix,BOONDOX-calo}
\usepackage{tocloft}
\usepackage{bbold}
\usepackage{tikz}
\usetikzlibrary{backgrounds}
\usetikzlibrary{patterns}
\usetikzlibrary{arrows.meta}
\usetikzlibrary{shapes,snakes}
\tikzstyle{bag} = [align=center]
\usetikzlibrary{decorations.pathmorphing}
\usepackage{hyperref}
\usepackage{subcaption}

 \newcommand{\badat}{\begin{alignedat}}
 \newcommand{\eadat}{\end{alignedat}}
 \newcommand\scalemath[2]{\scalebox{#1}{\mbox{\ensuremath{\displaystyle #2}}}}
 \def\be{\begin{equation}}
\def\ee{\end{equation}}

\def\p{\partial}

\usepackage{xcolor}

\newcommand{\pink}[1]{\textcolor{\pink}{#1}}

\definecolor{dblue}{rgb}{0.2,0.50,0.80}

 \definecolor{dred}{rgb}{0.65,0.10,0.20} 
\definecolor{dblue}{rgb}{0.2,0.50,0.80}

\def\N{\mathcal{N}}
\def\C{\mathcal{C}}
\def\O{\mathcal{O}}

\def\I{\mathcal{I}}

\def\Q{\mathcal{Q}}

\def\f{\mathcal{f}}

\def\bh{{\bar h}}
\def\bz{{\bar z}}
\def\bw{{\bar w}}

\def\tJ{\widetilde J}

\def\tN{\widetilde N}

\def\tDelta{\widetilde \Delta}
\def\tpsi{\widetilde \psi}

\def\tchi{\widetilde \chi}

\def\tbchi{\widetilde{\overline{\chi}}}
\def\bchi{\bar \chi}
\def\tbchi{\widetilde{\bar{\chi}}}

\def\tO{\widetilde \O}

\def\tDelta{\widetilde \Delta}

\def\tvarphi{\widetilde{\varphi}}

\def\sb{\bar{\sigma}}
\def\zb{\bar{z}}
\def\wb{\bar{w}}

\def\bh{{\bar h}}
\def\bz{{\bar z}}
\def\bw{{\bar w}}

\def\scri{\mathcal I}

\def\Gold{{\rm G}}

\numberwithin{equation}{section} 

\begin{document}

 \begin{titlepage}
  \thispagestyle{empty}
  \begin{flushright}
  CPHT-RR006.022021
  \end{flushright}
  \bigskip
  \begin{center}

        \baselineskip=13pt {\LARGE \scshape{
        Conformally Soft Fermions
        }}
  
      \vskip1cm 

   \centerline{ Yorgo Pano,${}^\blacklozenge{}$
   {Sabrina Pasterski},${}^\diamondsuit{}$
   {and Andrea Puhm}${}^\blacklozenge{}$
   }

\bigskip\bigskip

\centerline{\em${}^\blacklozenge$   CPHT, CNRS, Ecole Polytechnique, IP Paris, F-91128 Palaiseau, France}

\bigskip

 \centerline{\em${}^\diamondsuit$ Princeton Center for Theoretical Science, Princeton, NJ 08544, USA}

\smallskip

\bigskip\bigskip

\end{center}

\begin{abstract}
  \noindent 

Celestial diamonds encode the global conformal multiplets of the conformally soft sector, elucidating the role of soft theorems, symmetry generators and Goldstone modes.  Upon adding supersymmetry they stack into a pyramid. Here we treat the soft charges associated to the fermionic layers that tie this structure together. This extends the analysis of conformally soft currents for photons and gravitons which have been shown to generate asymptotic symmetries in gauge theory and gravity to infinite-dimensional fermionic symmetries. 
We construct fermionic charge operators in 2D celestial CFT from a suitable inner product between 4D bulk field operators and spin $s=\frac{1}{2}$ and~$\frac{3}{2}$ conformal primary wavefunctions with definite SL(2,$\mathbb{C}$) conformal dimension~$\Delta$ and spin~$J$ where $|J|\leq s$. The generator for large supersymmetry transformations is identified as the conformally soft gravitino primary operator with $\Delta=\frac{1}{2}$ and its shadow with $\Delta=\frac{3}{2}$ which form the left and right corners of the celestial gravitino diamond. We continue this analysis to the subleading soft gravitino and soft photino which are captured by degenerate celestial diamonds. Despite the absence of a gauge symmetry in these cases, they give rise to conformally soft factorization theorems in celestial amplitudes and complete the celestial pyramid.

\end{abstract}

\end{titlepage}

\tableofcontents

\newpage
\section{Introduction}

Understanding the symmetries of nature is a fundamental open question.  The fact that we can shed new light on this subject by reorganizing how we study scattering processes is central to the recently conjectured duality between quantum gravity in asymptotically flat spacetimes and a celestial conformal field theory living on the codimension-two sphere at null infinity.

Hints for such a duality were uncovered by studying infrared aspects of gauge theory and gravity.  A key insight~\cite{Strominger:2017zoo} is the fact that asymptotic symmetries not only present an infinite dimensional enhancement of the familiar global symmetries, but also are realized perturbatively in quantum field theory via soft theorems~\cite{Weinberg:1965nx}. 
However, the real power of this framework comes from pushing the correspondence to involve a full quantum theory of gravity. This is highlighted by the fact that the holographic map mixes infrared and ultraviolet limits of bulk observables. 
In this framework, $\mathcal{S}$-matrix elements get mapped to conformal correlators~\cite{Pasterski:2016qvg,Pasterski:2017kqt,Pasterski:2017ylz,deBoer:2003vf,Cheung:2016iub} by transforming the external states to boost, rather than energy, eigenstates. For the massless case, the map is simply a Mellin transform in the energies of each external scattering state
\be
\langle O^\pm_{\Delta_1}(z_1,\bz_1)...O^\pm_{\Delta_n}(z_n,\bz_n)\rangle=\prod_{i=1}^n \int_0^\infty d\omega_i \omega_i^{\Delta_i-1} \langle out|\mathcal{S}|in\rangle\,,
\ee
which trades the energy for a conformal dimension $\Delta$ while the 4D helicity gives the 2D spin $J$.
It is thus necessary to incorporate ultraviolet physics in order for basic observables like the gravitational $\mathcal{S}$-matrix to be well-defined in the new basis, making it natural to expect that string theory plays an important role.

For instance, when recast as a celestial correlator, the scattering amplitude for four radiative gravitons diverges. As shown in~\cite{Stieberger:2018edy}, this divergent behavior is softened in string theory. Moreover, the worldsheet of the string is pinned to the celestial CFT in certain limits.
Another recent insight is that the holographic symmetries of the MHV sector of gravitational scattering involves a ${\rm w}_{1+\infty}$ algebra hinting at a potential connection to $\N=2$ string theory~\cite{Strominger:2021lvk}. A full understanding of the relation between the string worldsheet CFT and celestial CFT, and the embedding of celestial holography into string theory are open problems.

 An essential step in building our holographic dictionary is to understand the spectrum and the symmetries of celestial CFT. These questions go hand in hand. 
 Data on the principal series $\Delta\in 1+i\mathbb{R}$ form a basis for finite energy states~\cite{Pasterski:2017kqt}.  However, a special role is played by $\Delta\in\frac{1}{2}\mathbb{Z}$, since soft theorems in momentum space map to conformally soft theorems at these values of the conformal dimension~\cite{Cheung:2016iub}.  
Each Goldstone/Goldstino of a spontaneously broken asymptotic symmetry corresponds to a conformally soft factorization theorem. This has been shown for gauge theory~\cite{Fan:2019emx,Nandan:2019jas,Pate:2019mfs} and gravity~\cite{Adamo:2019ipt,Puhm:2019zbl,Guevara:2019ypd} where the asymptotic symmetries are a large U(1) Kac-Moody symmetry~\cite{Strominger:2013lka,He:2014cra,He:2015zea} and BMS supertranslations~\cite{Strominger:2013jfa,He:2014laa} and superrotations~\cite{Kapec:2014zla,Kapec:2016jld,Himwich:2019qmj}. The corresponding symmetries are then generated by celestial currents constructed via suitable bulk inner products with conformal primary wavefunctions with the appropriate $\Delta$~\cite{Donnay:2018neh,Donnay:2020guq}.  For each of these symmetries, the conformal primary wavefunctions selecting the currents is pure gauge.

The question of whether or not these modes with real (half-)integer conformal dimension $\Delta$ augment the conformal basis on the principal continuous series of the SL(2,$\mathbb{C}$) Lorentz group~\cite{Pasterski:2017kqt} was addressed in~\cite{Donnay:2020guq}, where we showed how one can analytically continue the external states to these conformal dimensions. The freedom to do so not only lets us explore conformally soft dimensions, but also has the effect of rendering perturbative celestial amplitudes for non-scale invariant theories well defined in a distributional sense~\cite{Puhm:2019zbl,Ball:2019atb}. With the full range of the complex plane to explore, it is possible to look for additional conformally soft theorems beyond those that can be traced to a (large) gauge symmetry origin. 

Indeed, there appear to be more soft factorization theorems than asymptotic gauge symmetries account for. In the plane wave basis, such examples appear at subleading orders in a Taylor expansion near zero energy compared to the ones that correspond to gauge symmetries.
These include the subleading soft photon/gluon theorem~\cite{Lysov:2014csa,Campiglia:2016hvg} and the subsubleading soft graviton theorem~\cite{Campiglia:2017dpg}, as well as possible scalar asymptotic symmetries~\cite{Campiglia:2017dpg}. In supersymmetric theories, in addition to the soft gravitino which has an interpretation as large supersymmetry~\cite{Awada:1985by,Lysov:2015jrs,Avery:2015iix}, there exist soft gluino and subleading soft gravitino theorems - see e.g.~\cite{Liu:2014vva,Rao:2014zaa,Chen:2014xoa,Dumitrescu:2015fej}. In the conformal basis, all of these modes are nicely captured by celestial diamonds~\cite{Pasterski:2021fjn,Pasterski:2021dqe}.

For special values of the conformal dimension celestial primary operators will have SL(2,$\mathbb{C}$) primary descendants~\cite{Penedones:2015aga,Banerjee:2019aoy,Banerjee:2019tam}. Celestial diamonds~\cite{Pasterski:2021fjn,Pasterski:2021dqe} provide a celestial CFT interpretation for the structure of these conformal multiplets. As observed in~\cite{Pasterski:2021fjn}, this structure persists beyond the conformally soft theorems with asymptotic symmetry interpretations, capturing both the most subleading soft theorems, as well as the infinite towers of primaries giving symmetry enhancements in the single helicity sector~\cite{Guevara:2021abz,Strominger:2021lvk}.

Infinite-dimensional fermionic symmetries have received much less attention in the celestial CFT context than their bosonic counterparts~\cite{Donnay:2018neh,Donnay:2020guq,Pasterski:2021dqe,Banerjee:2019aoy,Banerjee:2019tam,Banerjee:2020zlg}. Moreover, subleading conformally soft theorems which do not have an obvious gauge symmetry origin have also remained elusive~\cite{Lysov:2014csa,Campiglia:2016hvg,Campiglia:2016jdj,Campiglia:2016efb,Himwich:2019dug,Banerjee:2021cly}. The purpose of this paper is to remedy this. Building off the fermionic soft theorems of~\cite{Dumitrescu:2015fej} for photinos and~\cite{Lysov:2015jrs,Avery:2015iix} for the gravitino, the conformal primary fermions of~\cite{Muck:2020wtx,Narayanan:2020amh,Pasterski:2020pdk}, as well as the results on conformally soft theorems in $\N=1$ supergravity by~\cite{Fotopoulos:2020bqj}, we  will see that celestial diamonds naturally provide a footing to study the conformally soft sector of celestial CFT when supersymmetry is incorporated.

This paper examines fermionic symmetries and the associated celestial diamonds relevant to supergravity and supersymmetric gauge theories. We identify the generator for large supersymmetry transformations as the conformally soft gravitino primary operator with SL(2,$\mathbb{C}$) dimension 
$\Delta=\frac{1}{2}$ and its shadow with conformal dimension 
$\Delta=\frac{3}{2}$ which form the left and right corners of the celestial gravitino diamond.\footnote{Note that while these values of the conformal dimension have already been identified in~\cite{Fotopoulos:2020bqj} from conformally soft factorization, here we explicitly construct the 2D operators as bona fide conformal primaries.} The structure of the diamond is chiral and only one of the two corners yields a Ward identity that is isomorphic to the conformally soft gravitino theorem. The primary descendant operator at the bottom corner of the gravitino diamond corresponds to the fermionic analogue of the soft charge operators of~\cite{Banerjee:2018fgd,Banerjee:2019aoy,Banerjee:2019tam}.

Since the formalism set up in~\cite{Pasterski:2020pdk} extends naturally to any spin, we take advantage of this to identify conformal primary operators of spin-$\frac{1}{2}$ and spin-$\frac{3}{2}$ which correspond to known soft theorems but which do not have a gauge symmetry origin. This includes the subleading soft gravitino and soft photino which correspond to degenerate celestial diamonds. Despite the absence of a gauge symmetry in these cases we can construct fermionic boundary operators that generate a shift in bulk fields akin to those generated by spontaneously broken asymptotic symmetries. For the conformally soft photino with $\Delta=\frac{1}{2}$ 
our 2D operator is consistent with the soft charge considered in~\cite{Dumitrescu:2015fej}. Meanwhile, a spacetime interpretation for the subleading soft gravitino has not been considered before in the literature; we fill this gap by computing a soft charge for the $\Delta=-\frac{1}{2}$ conformally soft gravitino
. These operators are further shown to select modes corresponding to the fermionic conformally soft theorems considered in~\cite{Fotopoulos:2020bqj}.
Because their existence is tied to supersymmetry~\cite{Fotopoulos:2020bqj,Pasterski:2020pdk} and double copy relations~\cite{Casali:2020vuy,Pasterski:2020pdk} which connect fields of lower spin to those with large gauge symmetries, we expect them to play an important role in understanding the analog of the spin shifting amplitudes relations within celestial CFT. Indeed, we see that these fermionic and bosonic diamonds glue together into a celestial pyramid, which nicely unites the spin-shifting analysis of~\cite{Pasterski:2020pdk} and conformal multiplet studies of~\cite{Pasterski:2021dqe,Pasterski:2021fjn}.

This paper is organized as follows.  We set up the conformal primary wavefunctions and operators for celestial fermions in section~\ref{sec:celestialfermions}.  We then examine the behavior of the bulk fields near null infinity in section~\ref{sec:BulkBoundary}  to establish the relevant extrapolate dictionary. With this technology we are ready to examine the Goldstino mode corresponding to the leading soft graviton theorem in section~\ref{sec:gravitino}.  We then use the structure of celestial diamonds to extend these constructions to fermions with conformally soft theorems but no corresponding conformal Goldstino modes in section~\ref{sec:subgoldstinos}. We conclude by examining how supersymmetry  relates the fermionic and bosonic examples in section~\ref{celestialpyramids}. Further computational details can be found in the appendix.

\section{Celestial Fermions}
\label{sec:celestialfermions}

In this section, we set up the conformal primary wavefunctions for spin-$\frac{1}{2}$ and spin-$\frac{3}{2}$ fields and use these to construct the celestial operators corresponding to single particle scattering states.

\subsection{Conformal Primary Wavefunctions}\label{sec:CPW}
{\it Conformal primary wavefunctions} are functions on $\mathbb{R}^{1,3}\times \mathbb{C}$ which depend on a spacetime vector $X^\mu\in \mathbb{R}^{1,3}$ and a point $(w,\bw) \in S^2$, which under simultaneous SO(1,3)$\simeq$ SL(2,$\mathbb{C}$) Lorentz transformations of $X$ and M\"{o}bius transformation of $(w,\bw)$ on the celestial sphere
 \be\label{mobius}
X^\mu\mapsto \Lambda^\mu_{~\nu}X^\nu\,,~~~ w\mapsto \frac{a w+b}{cw+d}\,,~~~\bw\mapsto \frac{{\bar a} \bw+{\bar b}}{{\bar c}\bw+{\bar d}}\,,
 \ee
transform as 2D conformal primaries with SL(2,$\mathbb{C}$) conformal dimension~$\Delta$ and 2D spin~$J$. 
We focus here on fermionic wavefunctions and consider two types: {\it radiative} conformal primary wavefunctions which satisfy the equations of motion for massless spin-$s$ fields in the vacuum and have $J=\pm s$, and {\it generalized} conformal primary wavefunctions with $|J|\leq s$ and in principle allow for sources and non-analytic behavior though here we restrict ourselves to analytic wavefunctions.

\subsubsection*{A Covariant Tetrad and Spin Frame}
Let us first introduce some useful notation.  We start by embedding the celestial sphere into the $\mathbb{R}^{1,3}$ lightcone via the null reference direction
\be\label{qmu}
q^\mu=(1+w\bw,w+\bw,i(\bw-w),1-w\bw)\,,
\ee
from which we construct the following null tetrad for Minkowski space~\cite{Pasterski:2020pdk} 
\be\label{tetrad}
l^\mu=\frac{q^\mu}{-q\cdot X}\,, ~~~n^\mu=X^\mu+\frac{X^2}{2}l^\mu\,, ~~~m^\mu=\epsilon^\mu_++(\epsilon_+\cdot X) l^\mu\,, ~~~\bar{m}^\mu=\epsilon^\mu_-+(\epsilon_-\cdot X) l^\mu\,,
\ee
 where $\epsilon_+^\mu=\frac{1}{\sqrt{2}}\p_w q^\mu$ and $\epsilon_-^\mu=\frac{1}{\sqrt{2}}\p_{\bw} q^\mu$.
 The vectors of this tetrad  satisfy standard normalization conditions $l\cdot n=-1\,,~~m\cdot\bar{m}=1$ with all other inner products vanishing and have the property that they transform covariantly under the SL(2,$\mathbb{C}$)
\be\label{tetradcov}
l^\mu\mapsto \Lambda^{\mu}_{~\nu} l^\nu\,,~~n^\mu\mapsto \Lambda^{\mu}_{~\nu} n^\nu\,,
~~m^\mu\mapsto \frac{cw+d}{{\bar c}\bw+{\bar d}}\Lambda^{\mu}_{~\nu} m^\nu\,,~~{\bar m}^\mu\mapsto \frac{{\bar c}\bw+{\bar d}}{cw+d}\Lambda^{\mu}_{~\nu} {\bar m}^\nu\,,
\ee
where $\Lambda^{\mu}_{~\nu}$ is the corresponding vector representation of SO(1,3)$\simeq$ SL(2,$\mathbb{C}$).
 To discuss operators and wavefunctions of half-integer spin it is convenient to further decompose the tetrad into a spin frame
\be\label{spinframe}
l_{a{\dot b}}=o_a\bar{o}_{\dot b}\,,~~n_{a{\dot b}}=\iota_a\bar{\iota}_{\dot b}\,,~~m_{a{\dot b}}=o_a{\bar\iota}_{\dot b}\,,~~\bar{m}_{a{\dot b}}=\iota_a{\bar o}_{\dot b}\,,
\ee
where $v_{a{\dot b}}=v_\mu\sigma^\mu_{a\dot b}$ for $\sigma^\mu_{a\dot b}=(\mathds{1},\sigma^i)_{a\dot b}$ with the Pauli matrices $\sigma^i$. This yields
\be\label{spinframespinors}
o_a=\sqrt{\frac{2}{q\cdot X}}\left(\begin{array}{c} \bw \\-1 \end{array}\right)\,,~~~\iota_a=\sqrt{\frac{1}{q\cdot X}}\left(\begin{array}{c}X^0-X^3-w(X^1-iX^2) \\-X^1-iX^2+w(X^0+X^3) \end{array}\right)\,,
\ee
up to an overall phase ambiguity which is fixed by setting ${\bar o}_{\dot a}=(o_a)^*$ and ${\bar \iota}_{\dot a}=(\iota_a)^*$ in the region where $q\cdot X>0$ and analytically continued from there. The elements of the spin frame transform as 
\be
o_a\mapsto (cw+d)^{\frac{1}{2}}({\bar c}\bw+{\bar d})^{-\frac{1}{2}}(Mo)_a\,, \quad \iota_a\mapsto (cw+d)^{-\frac{1}{2}}({\bar c}\bw+{\bar d})^{\frac{1}{2}}(M\iota)_a\,,
\ee
where $M$ is an element of $\overline{ SL(2,\mathbb{C})}$.  We refer to appendix~\ref{app:Conventions} for our spinor conventions.

\subsubsection*{Radiative Conformal Primary Wavefunctions}
The scalar conformal primary wavefunction corresponds to the Mellin transform of a plane wave,
\begin{equation}\label{varphi}
    \varphi^{\Delta,\pm} =\frac{1}{(-q\cdot X_\pm)^\Delta}=\frac{1}{(\mp i)^\Delta \Gamma(\Delta)}\int_0^\infty d\omega \omega^{\Delta-1} e^{\pm i \omega q\cdot X- \varepsilon q^0\omega}\,,
\end{equation}
where $X^\mu_\pm=X^\mu\pm i\varepsilon\{-1,0,0,0\}$ is used as a regulator. Except as necessary, we will omit the $\pm$~label henceforth. In terms of~\eqref{varphi} together with the null tetrad~\eqref{tetrad} and the spin frame~\eqref{spinframe} the radiative spin-$\frac{1}{2}$ and spin-$\frac{3}{2}$ wavefunctions are given by~\cite{Pasterski:2020pdk}
\be\begin{array}{ll}\label{psichi}
\psi_{\Delta,J=+\frac{1}{2}}=o\varphi^\Delta\,,&~~~{\bar\psi}_{\Delta,J=-\frac{1}{2}}={\bar o}\varphi^\Delta\,,\\
\chi_{\Delta,J=+\frac{3}{2};\mu}=m_\mu o\varphi^\Delta\,,&~~~{\bar\chi}_{\Delta,J=-\frac{3}{2};\mu}={\bar m}_\mu{\bar o}\varphi^\Delta\,,
\end{array}
\ee
with the expressions on the left denoting left-handed spinors, while the ones on the right correspond to right-handed spinors. To simplify the notation we have omitted the spinor indices.

Related to the conformal primaries~\eqref{varphi}
-\eqref{psichi} via a shadow transform are wavefunctions of flipped and shifted conformal dimension and flipped spin.
The scalar shadow wavefunction is given by
\begin{equation}\label{SHvarphi}
    {\tvarphi}^{\Delta}=(-X^2)^{\Delta-1}\varphi^{\Delta}\,,
\end{equation}
while we claimed in~\cite{Pasterski:2020pdk} that the shadow wavefunctions for spin-$\frac{1}{2}$ and spin-$\frac{3}{2}$ are given by\footnote{We have chosen the normalization conventions of~\cite{Pasterski:2021fjn} which differ from those of~\cite{Pasterski:2017kqt,Pasterski:2020pdk} by an overall sign in the $s=\frac{1}{2}$ shadow wavefunctions.} 
\be\begin{array}{ll}\label{SHpsichi}
{\tpsi}_{\Delta,J=-\frac{1}{2}}=-\sqrt{2}\iota(-X^2)^{\Delta-\frac{3}{2}}\varphi^\Delta\,,&~~~\widetilde{{\bar\psi}}_{\Delta,J=+\frac{1}{2}}=-\sqrt{2}{\bar \iota}(-X^2)^{\Delta-\frac{3}{2}}\varphi^\Delta\,,\\
{\tchi}_{\Delta,J=-\frac{3}{2};\mu}=\sqrt{2}{\bar m}_\mu \iota(-X^2)^{\Delta-\frac{3}{2}}\varphi^\Delta\,,&~~~\widetilde{{\bar\chi}}_{\Delta,J=+\frac{3}{2};\mu}=\sqrt{2}{ m}_\mu{\bar \iota}(-X^2)^{\Delta-\frac{3}{2}}\varphi^\Delta\,.\\
\end{array}
\ee
We will explicitly verify the expressions~\eqref{SHpsichi} in an expansion near null infinity in appendix~\ref{app:radexpCPW}.

\subsubsection*{Generalized Conformal Primary Wavefunctions}
Generalized bulk wavefunctions for half-integer spins $s\leq |J|$ were discussed in~\cite{Pasterski:2020pdk,Pasterski:2021fjn}. We focus here on the analytic wavefunctions built from the generalized scalar $\varphi^{gen}_\Delta=f(X^2) \varphi^\Delta$. For $s=\frac{1}{2}$ they are given by
\begin{equation}
    \psi^{gen}_{\Delta,+\frac{1}{2}}=o \varphi^{gen}_\Delta\,, \quad \psi^{gen}_{\Delta,-\frac{1}{2}}=\iota \varphi^{gen}_\Delta\,.
\end{equation}
Enforcing the vacuum Weyl equation $\bar{\sigma}^\mu\p_\mu \psi^{gen}_{\Delta,\pm\frac{1}{2}}=0$ yields the radiative conformal primary wavefunctions~\eqref{psichi} and their shadows~\eqref{SHpsichi} for generic $\Delta \in \mathbb{C}$ and $J=\pm \frac{1}{2}$. For $s=\frac{3}{2}$ they are given by 
\begin{equation}
    \chi^{gen}_{\Delta,+\frac{3}{2};\mu}=m_\mu o\varphi_{\Delta}^{gen}\,, \quad \chi^{gen}_{\Delta,-\frac{3}{2};\mu}=\bar{m}_\mu\iota\varphi_{\Delta}^{gen}\,,
\end{equation}
and
\begin{equation}
    \chi^{gen}_{\Delta,+\frac{1}{2};\mu}=l_\mu o\varphi_{\Delta}^{gen,1}+n_\mu o\varphi_{\Delta}^{gen,2}+m_\mu \iota\varphi_{\Delta}^{gen,3}\,, \quad \chi^{gen}_{\Delta,-\frac{1}{2};\mu}=l_\mu \iota\varphi_{\Delta}^{gen,1}+n_\mu \iota\varphi_{\Delta}^{gen,2}+\bar{m}_\mu o\varphi_{\Delta}^{gen,3}\,,
\end{equation}
where $\varphi^{gen,i}_\Delta=f_i(X^2) \varphi^\Delta$.
Enforcing the chiral projection of the Rarita-Schwinger equation $\varepsilon^{\mu\nu\rho\kappa}\bar{\sigma}_\nu\nabla_\rho \chi^{gen}_{\Delta,J;\kappa}=0$ as well as the gauge conditions
\be \nabla^\mu \chi^{gen}_{\Delta,J;\mu}=0\,,~~ X^\mu \chi^{gen}_{\Delta,J;\mu}=0\,,~~ \sb^\mu \chi^{gen}_{\Delta,J;\mu}=0\,,\ee
yields the radiative conformal primary wavefunctions~\eqref{psichi} and their shadows~\eqref{SHpsichi} for generic $\Delta \in \mathbb{C}$ and $J=\pm \frac{3}{2}$, as well as a discrete set of solutions $\Delta=\frac{5}{2}$ and $J=\pm \frac{1}{2}$ which we will come back to in section~\ref{sec:gravitino}.

\subsection{Conformal Primary Operators}\label{sec:CPO}

Given a 4D operator of spin-$s$ in the Heisenberg picture, we can define a 2D celestial CFT operator via a suitable inner product with the conformal primary wavefunctions. We need three ingredients: our primary wavefunctions, a bulk operator, and an appropriate inner product. We examined spinorial primary wavefunctions in the previous section, so let us move on to the operator mode expansions.

\paragraph{Bulk Operator Mode Expansions}
In the momentum basis, we have the following mode expansions for the spin-$\frac{1}{2}$
and spin-$\frac{3}{2}$ left-handed Weyl components of the Majorana fields. 
Parametrizing null momenta as $k^\mu=\omega q^\mu$ we can express them as a product of spinor-helicity variables
\begin{equation}
    k_{a \dot a}=\sigma^\mu_{a\dot a} k_\mu = -|k]_a \langle k |_{\dot a}\,, 
\end{equation}
where
\begin{equation}
    |k]_a=\sqrt{\omega}|q]_a\,, \quad \langle k|_{\dot a}=\sqrt{\omega}\langle q|_{\dot a}\,,
\end{equation}
with
\be
|q]_a=\sqrt{q\cdot X}o_a~~~\langle q|_{\dot{a}}=\sqrt{q\cdot X}\bar{o}_{\dot{a}}\,.
\ee
For the left-handed massless Weyl photino we have the mode expansion
\begin{equation}\label{hatpsiBulk}
    \hat \psi_a(X)=e\int \frac{d^3k}{(2\pi)^3}\frac{|k]_a}{2k^0}\left[a_- e^{ik\cdot X}+a_+^\dagger e^{-ik\cdot X}\right]\,, 
\end{equation}
while for the left-handed massless Weyl gravitino we have 
\begin{equation}\label{hatchiBulk}
   \hat{\chi}_{\mu a}(X)=\kappa\int \frac{d^3k}{(2\pi)^3}\frac{|k]_a}{2k^0}\epsilon^+_\mu\left[ a_- e^{ik\cdot X}+a_+^\dagger e^{-ik\cdot X}\right]\,.
\end{equation}
These mode operators obey the standard anti-commutation relations
\be
\{a_\pm(k),a^\dagger_\pm(k')\}=(2\pi)^3 (2k^0)\delta^{(3)}(k-k')\,.
\ee
The opposite helicity modes are captured by the right-handed Weyl spinors.

\paragraph{From Weyl to Dirac and Majorana Primaries}
The  spin $s=\frac{1}{2}$ and $\frac{3}{2}$ Weyl spinors corresponding to the photino and gravitino can be embedded into Dirac spinors 
\be\label{embed1}
\Psi^{s=\frac{1}{2}}=\left(\begin{array}{c}
     \psi_a  \\ \bar{\psi}^{\dot a}
\end{array}\right)\,, \quad \Psi^{s=\frac{3}{2}}_{\mu}=\left(\begin{array}{c}
     \chi_{\mu a}  \\ \bar{\chi}_{\mu}^{\dot a}
\end{array}\right)\,,
\ee
on which we impose a Majorana condition (see appendix~\ref{app:Conventions} for more details)
\begin{equation}\label{majorana}
    \bar{\psi}^{\dot a}=\varepsilon^{\dot a \dot b}(\psi^\dagger)_{\dot b}\,, \quad  \bar{\chi}^{\dot a}_\mu=\varepsilon^{\dot a \dot b}(\chi^\dagger)_{\mu\dot b}\,.
\end{equation}
The conformal primary wavefunctions of section~\ref{sec:CPW} embed into Dirac spinors of definite conformal dimension and spin
\be\label{chiralspinor}
\Psi_{\Delta,J}^{s=\frac{1}{2}}=\left(
\begin{array}{c}
 \psi_{\Delta,J} \\
  \bar{\psi}_{\Delta,J}
\end{array}\\ \right)\,, \quad 
\Psi_{\Delta,J;\mu}^{s=\frac{3}{2}}=\left(
\begin{array}{c}
 \chi_{\Delta,J;\mu} \\
  \bar{\chi}_{\Delta,J;\mu}
\end{array}\\ \right)\,,
\ee
where the component spinors are given by~\eqref{psichi}. Thus for fixed sign of the spin only one Weyl component is non-vanishing.  Because these primaries obey
\be
(\Psi_{\Delta,J})^C=\Psi_{\Delta^*,-J}\,,
\ee
where the charge conjugate $\Psi^C$ is defined in~\eqref{chargeconjugate}, we can construct a Majorana primary for either $s=\frac{1}{2}$ and $\frac{3}{2}$ (omitting the spacetime index) as
\be\label{embed}
\Psi_{\rm M}(\Delta,J)=\Psi_{\Delta,J}+\Psi_{\Delta^*,-J}\,.
\ee

\subsubsection*{Fermionic 2D Operators}

With these ingredients, as well as the inner products which we review in appendix~\ref{app:IP}, we are now ready to construct fermionic primary operators generalizing the bosonic construction of~\cite{Donnay:2020guq,Pasterski:2021fjn}.  In terms of the Dirac spinors~\eqref{chiralspinor} we define the 2D fermionic operator
\be\label{Ffermionic}
\mathcal{O}^{s,\pm}_{\Delta,J}(w,\bw)\equiv i(\hat{\Psi}^s(X^\mu),\Psi^s_{\Delta^*,-J}(X_\mp^\mu;w,\bw))\,,
\ee
where $\pm$ on the operator indicates whether it corresponds to an {\it in} or an {\it out} state. The inner products $(.\,,.)$ for $s=\frac{1}{2}$ and $\frac{3}{2}$ Dirac spinors are given in~\eqref{IPphotino} and~\eqref{IPgravitino}, respectively.  In terms of our Weyl spinors these take the explicit form 

\begin{equation}\label{Fphotinos}
\badat{2}
    \O^{s,\pm}_{\Delta,J}(w,\bw)&=i\int d\Sigma_\nu \left({\psi}_{\Delta,J} (X_\pm;w,\bw) \sigma^\nu  \hat {{\psi}}^\dagger(X)+\bar{\psi}_{\Delta,J} (X_\pm;w,\bw) \bar\sigma^\nu  \hat {\psi}(X)\right)\,,
\eadat
\end{equation}
for the photino and
\begin{equation}\label{Fgravitinos}
\badat{2}
    \O^{s,\pm}_{\Delta,J}(w,\bw)&=i\int d\Sigma_\nu \left(\chi_{\Delta,J;\mu} (X_\pm;w,\bw) \sigma^\nu  \hat \chi^{\dagger\mu}(X)+ \bar{\chi}_{\Delta,J;\mu}(X_\pm;w,\bw)\bar\sigma^\nu \hat{\chi}^\mu(X) \right)\,,
\eadat
\end{equation}
for the gravitino, where we note that for fixed $\Delta,J$ only one of the two terms in each expression will be non-zero. We will use $\widetilde{\mathcal{O}}^{s,\pm}_{\Delta,J}$ to denote the shadow operator constructed from~\eqref{Ffermionic} with~\eqref{SHpsichi} replacing~\eqref{psichi} in the chiral Dirac spinor~\eqref{chiralspinor}.

For integer spins $s=1,2$ we showed in~\cite{Donnay:2020guq} that for certain values of $\Delta$ the operators $\O^s_{\Delta,J}(w,\bw)$ correspond to soft charges in the full (matter-coupled) theory when the Cauchy slice on which they are defined is taken to null infinity and the wavefunctions $\Phi^s_{\Delta,J}$ are the Goldstone modes of the spontaneously broken asymptotic symmetries in gauge theory and gravity.  Here, the operator~\eqref{Ffermionic} generates the shift\footnote{Recall that the mode operators are Grassmann valued while our conventions are such that the fermionic primary wavefunctions are not, as such one should multiply these generators by a Grassmann valued parameter to implement the supersymmetry transformations. See~\cite{Fotopoulos:2020bqj,Jiang:2021xzy,Brandhuber:2021nez,Hu:2021lrx} for recent work on celestial superfields.}
\begin{equation}\label{shift2}
\{\O^{s,\pm}_{\Delta,J}(w,\bw),\hat \Psi^s(X)\}=i\Psi^s_{\Delta,J}(X_\mp;w,\bw)\,.
\end{equation}
In section~\ref{sec:gravitino} we will show that the spin-$\frac{3}{2}$ operator~\eqref{Ffermionic} for certain values of $\Delta$ corresponds to the soft charge for spontaneously broken large supersymmetry whose Ward identity is equivalent to the conformally soft gravitino theorem. Moreover, even in the absence of an asymptotic symmetry, we will identify~\eqref{Ffermionic} for $s=\frac{1}{2},\frac{3}{2}$ with soft charges associated to the conformally soft photino theorem and the subleading conformally soft gravitino theorem.

\section{From Bulk to Boundary}\label{sec:BulkBoundary}

Evaluating the fermionic 2D operators at null infinity requires knowing both the large-$r$ expansions of the 4D field operators and the conformal primary wavefunctions.
Near $\scri^+$ we use retarded Bondi coordinates $(u,r,z,\bz)$ which are related to the Cartesian coordinates $(X^0,X^1,X^2,X^3)$ by the transformation
\begin{equation}\label{eq:co}
 X^0=u+r\,, \quad X^i=r \hat{X}^i(z,\bz)\,, \quad \hat{X}^i(z,\bz)=\frac{1}{1+z \bar z}(z+\bz,i(\bz-z),1-z\bz)\,,
\end{equation}
which maps the line element to
\begin{equation}\label{gBondi}
ds^2=-du^2-2du dr+2r^2 \gamma_{z \bar z} dz d\bar z \quad \text{with}\quad \gamma_{z \bar z}=\frac{2}{(1+z \bar z)^2}\,.
\end{equation} 
To discuss spinor fields we use the flat frame
\begin{equation}
e^0=e^0_\mu dX^\mu\equiv du+dr\,, \quad e^i=e^i_\mu dX^\mu \equiv \hat X^idr+r\partial_z \hat X^i dz+r\partial_\bz \hat X^i d\bz\,,
\end{equation}
for which the spin connection vanishes. The Gamma matrices in retarded Bondi coordinates are of the form $\gamma_\mu\equiv e_\mu^0 \gamma_0+e_\mu^i \gamma_i$ yielding
\begin{equation}
    \gamma_u =\gamma_0\,, \quad \gamma_r=\gamma_0+\hat X^i \gamma_i\,, \quad \gamma_z=r\partial_z \hat X^i \gamma_i\,, \quad \gamma_\bz=r\partial_\bz \hat X^i \gamma_i\,.
\end{equation}

\subsection{Operator Expansion}\label{sec:OPexpansion}
At the center of the equivalence between soft theorems and Ward identities of asymptotic symmetries lies the
large-$r$ saddle point approximation
\be\label{saddlept}
\lim_{r\rightarrow\infty}\sin\theta e^{i\omega q^0r(1-\cos\theta)}=\frac{i}{\omega q^0 r}\delta(\theta)+\mathcal{O}((\omega q^0 r)^{-2})\,,
\ee
which we use to evaluate the integrals~\eqref{hatpsiBulk}-\eqref{hatchiBulk} and their Hermitian conjugates.
For spin-$\frac{1}{2}$ we have
\be\label{psiScri}
\lim\limits_{r\rightarrow\infty} r\hat{\psi}=-\frac{i e}{2(2\pi)^2} (1+z\bz) \int_0^\infty d\omega {\omega}^{\frac{1}{2}}\left[ a_-(\omega,z,\bz)e^{-i\omega (1+z\bz) u} -  a^\dagger_+(\omega,z,\bz)e^{i\omega (1+z\bz) u}\right]|x]\,,
\ee
while for spin-$\frac{3}{2}$ we get
\be\label{chiScri}
\lim\limits_{r\rightarrow\infty} \hat{\chi}_\bz=-\frac{i \kappa}{\sqrt{2}(2\pi)^2} \int_0^\infty d\omega {\omega}^{\frac{1}{2}}\left[ a_-(\omega,z,\bz)e^{-i\omega (1+z\bz) u} -  a^\dagger_+(\omega,z,\bz)e^{i\omega (1+z\bz) u}\right]|x]\,,
\ee
with the other components subleading. Replacing $a_- \mapsto a_+$, $a^\dagger_+\mapsto a^\dagger_-$ and $|x]_a \mapsto |x\rangle^{\dot a}$ yields the Hermitian conjugate Weyl spinors $\hat{\psi}^\dagger$ and $\hat{\chi}^\dagger_z$ where we note that $ \epsilon^+_\bz=\epsilon^-_z=\frac{\sqrt{2}r}{1+z\bz}$ while $\epsilon^+_z= \epsilon^-_\bz=0$.
The saddle point approximation has localized $(w,\bw)\mapsto (z,\bz)$ and we have introduced 
\begin{equation}
    |x]_a=\sqrt{2}\left(\begin{array}{c} \bz\\-1\end{array}\right)\,, \quad  | x\rangle^{\dot a}=-\sqrt{2}\left(\begin{array}{c} 1\\z\end{array}\right).\,
\end{equation}

The above expressions~\eqref{psiScri}-\eqref{chiScri} give the bulk operators near null infinity using a momentum basis mode expansion. The creation and annihilation operators with definite $(\Delta,J)$ are obtained from $a_\pm$ and $a^\dagger_\pm$ via a Mellin transform
\be\label{mellinmode}
a_{\Delta,\pm s}=\int_0^\infty d\omega \omega^{\Delta-1}a_\pm(\omega)\,,\quad a^\dagger_{\Delta,\mp s}=\int_0^\infty d\omega \omega^{\Delta-1}a^\dagger_\pm(\omega)\,,
\ee
while the inverse Mellin transform
\begin{equation}\label{inversemellinmode}
    a_\pm(\omega)=\frac{1}{2\pi} \int_{1-i\infty}^{1+i\infty} (-id\Delta) \omega^{-\Delta}a_{\Delta,\pm s}\,,\quad    a^\dagger_\mp(\omega)=\frac{1}{2\pi} \int_{1-i\infty}^{1+i\infty} (-id\Delta) \omega^{-\Delta}a^\dagger_{\Delta,\pm s}\,,
\end{equation}
takes us back to the momentum basis. We have labeled both the creation and annihilation Mellin operators with the $\Delta$ and $J$ of the corresponding state on the celestial sphere.

\subsection{Wavefunction Expansion}\label{sec:CPWexpansion}

Finally, we need the expansion of the conformal primary wavefunctions near null infinity in order to evaluate the 2D operators and identify them for special values of $\Delta$ with soft charges. To obtain the large~$r$ expansion for the fermions we note that the components of the spin frame can be written in terms of the scalar primary as
\begin{equation}\label{oiotaexpansion}
    o=\sqrt{2}i\varphi^\frac{1}{2}\left( \begin{array}{c} \bw \\ -1  \end{array}\right)\,, \quad \iota=i\varphi^{\frac{1}{2}}\left[2r \frac{z-w}{1+z\bz} \left( \begin{array}{c} \bz \\ -1  \end{array}\right) +u \left( \begin{array}{c} 1 \\ w  \end{array}\right) \right]\,,
\end{equation}
while the relevant tetrad vector components can be expressed as
\be\label{mmbarexpansion}
m_{z}=-\sqrt{2}r(2r+u)\varphi^{1}\frac{(\bz-\bw)^2}{(1+z\bz)^2},~~~m_{\bz}=\sqrt{2}ru\varphi^{1}\frac{(1+z\bw)^2}{(1+z\bz)^2}\,,
\ee
with the expressions for $\bar{m}$ following from complex conjugation. The behavior near null infinity of the fermionic conformal primaries is thus dictated by the large-$r$ expansion of the scalar wavefunction. 
For the Mellin transformed plane wave the saddle-point approximation gives
  \begin{equation}\badat{3}\label{eq:sp}
  \lim_{r\rightarrow\infty} \int_0^\infty d\omega \omega^{\Delta-1}e^{\pm i\omega q\cdot X-\varepsilon\omega (1+z\bz)}
  &=r^{-1}\frac{u^{1-\Delta}\Gamma(\Delta-1)}{(\pm i)^{\Delta}(1+z\bz)^{\Delta-2}} \pi \delta^{(2)}(z-w)+\O(r^{-2})\,,
\eadat\end{equation}
which implies for the scalar primary
\begin{equation}
     \lim_{r\to \infty}\varphi^\Delta\Big|_{z= w}=r^{-1}\frac{u^{1-\Delta}}{\Delta-1} (1+z\bz)^{2-\Delta}\pi \delta^{(2)}(z-w)+\O(r^{-2})\,.
\end{equation}
Note that the $\omega$ integral in~\eqref{eq:sp} converges so long as Re$(\Delta)>1$, and we have taken the large~$r$ limit before integrating over~$\omega$. 
Going from the plane wave to the conformal basis first it is easy to see that away from $z=w$ we have the expansion
\begin{equation}
    \lim_{r\to \infty}\varphi^\Delta\Big|_{z\neq w}=\left(\frac{1+z\bz}{2(z-w)(\bz-\bw)}\right)^\Delta r^{-\Delta}+\O(r^{-\Delta-1})\,,
\end{equation}
which for Re$(\Delta)< 1$ is leading compared to the contact term~\eqref{eq:sp}.
For Re$(\Delta)= 1$ the above expressions contribute at the same order.\footnote{The appearance of similar ``resonances" were discussed e.g. in~\cite{Kutasov:1999xu}.} For the shadow transformed scalar primary~\eqref{SHvarphi} the roles of the contact and non-contact terms are reversed. Using $-X^2=u(2r+u)$ we find
\begin{equation}
   \lim_{r\to \infty}  \tvarphi^\Delta\Big|_{z= w}=r^{\Delta-2}\frac{2^{\Delta-1}}{\Delta-1}(1+z\bz)^{2-\Delta}\pi\delta^{(2)}(z-w)+\O(r^{\Delta-3})\,,
\end{equation}
with the $\omega$-integral~\eqref{eq:sp} again converging so long as Re$(\Delta)>1$, while
\begin{equation}
    \lim_{r\to \infty} \tvarphi^\Delta\Big|_{z\neq w}=r^{-1}u^{-1+\Delta} \frac{1}{2}\left(\frac{1+z\bz}{(z-w)(\bz-\bw)}\right)^\Delta +\O(r^{-2})\,.
\end{equation}
 Notice that at generic points ($z\neq w$) the shadow primaries have the standard radiative fall offs near null infinity, e.g. $\sim 1/r$ for scalar wavefunctions and this continues to hold for non-zero spin. Meanwhile the contact terms of the non-shadowed primaries are of radiative order.

The above ingredients determine the large-$r$ behavior of the photino and gravitino and their shadow transforms (see appendix~\ref{app:radexpCPW} for their detailed expressions). In particular, the photino arises from the scalar with conformal dimension shifted by $\frac{1}{2}$ while the gravitino is proportional to the scalar with conformal dimension shifted by $\frac{3}{2}$. 
In section~\ref{sec:gravitino} and~\ref{sec:subgoldstinos} we will be interested in the large-$r$ behavior for special half-integer values of the conformal dimension. For the leading conformally soft gravitino with $\Delta=\frac{1}{2}$ both contact and non-contact terms will appear at the same order and yield finite contributions, while for the conformally soft photino with $\Delta=\frac{1}{2}$ and the subleading conformally soft gravitino $\Delta=-\frac{1}{2}$ the contact terms have simple poles, respectively, $\frac{1}{\Delta-\frac{1}{2}}$ and $\frac{1}{\Delta+\frac{1}{2}}$.

\subsection{Extrapolate-Style Dictionary for Fermions}\label{sec:CPOexpansion}
We close this section by commenting on an extrapolate-style dictionary for fermions. Notice that states of definite conformal dimension are prepared with bulk operators integrated along a light ray of the form 
\be\label{lightray}
\mathcal{O}^{\frac{1}{2},-}_{\Delta,+\frac{1}{2}}\propto\lim_{r\rightarrow \infty} r\int du u_+^{-\Delta+\frac{1}{2}} \hat{\psi}(u,r,z,\bz), \qquad \mathcal{O}^{\frac{3}{2},-}_{\Delta,+\frac{3}{2}}\propto\lim_{r\rightarrow \infty} \int du u_+^{-\Delta+\frac{1}{2}} \hat{\chi}_{\bz}(u,r,z,\bz) \, ,
\ee
for the $+$ helicity modes, and similarly with the $z$ component for the $-$ helicity modes. 
Here we have defined $u_\pm=u\pm i \varepsilon$.  We note that a cancellation of phases selects the annihilation or creation operator in the integrals, namely
\be\scalemath{0.98}{\label{uplus}
\int_{-\infty}^\infty du u_+^{-\Delta+\frac{1}{2}} \int_0^\infty d\omega {\omega^{\frac{1}{2}}}\left[a_\pm e^{-i\omega(1+z\bz) {u}}-a^\dagger_{\mp}e^{i\omega(1+z\bz) {u}}\right]
=\frac{2\pi  (1+z\bz)^{\Delta-\frac{3}{2}}  }{{i^{\Delta-\frac{1}{2}}}\Gamma(\Delta-\frac{1}{2})} a_{\Delta,\pm s}\,,
}\ee
and 
\be\scalemath{0.98}{\label{uminus}
\int_{-\infty}^\infty du u_-^{-\Delta+\frac{1}{2}} \int_0^\infty d\omega{\omega^{\frac{1}{2}}} \left[a_\pm e^{-i\omega(1+z\bz) {u}}-a^\dagger_{\mp}e^{i\omega(1+z\bz) {u}}\right]={-}\frac{2\pi  (1+z\bz)^{\Delta-\frac{3}{2}} }{{(-i)^{\Delta-\frac{1}{2}}}\Gamma(\Delta-\frac{1}{2})}  a_{\Delta,\pm s}^\dagger
\,.}
\ee
Light ray operators at future (past) null infinity (with an appropriate analytic continuation off the real manifold) thus create the $\emph{out}$ ($\emph{in}$) states from the vacuum.
For example, the standard outgoing momentum eigenstates of a positive helicity photino is prepared via
\be\label{pout}
\langle p,+|{\propto}\lim\limits_{r\rightarrow\infty} r \int du\,  e^{i\omega(1+z\bz) u}\langle 0|\hat{\psi}\,, 
\ee
where $p_\mu=\omega q_\mu(z,\bz)$.  Here we are concerned with the state in the Hilbert space. The right hand side is also proportional to the spinor $|p]$, which we can remove by contracting with $\frac{1}{\omega(1+z\bz)}\langle p|\bar{\sigma}_u$.
Its analog for an outgoing boost eigenstate of SL(2,$\mathbb{C}$) spin $J=+\frac{1}{2}$ is 
 \be
\langle \Delta,z,\bz,+|{\propto}\lim_{r\rightarrow \infty} r \int du \,u_+^{-\Delta+\frac{1}{2}} \langle 0|\hat{\psi}(u_+,r,z,\bz)\, .
\ee
Similar expressions are obtained for the outgoing state of the gravitino with $J=+\frac{3}{2}$.

\section{Large Supersymmetry and Conformally Soft Gravitino}\label{sec:gravitino}
In this section we investigate the celestial diamond corresponding to the leading soft gravitino theorem and large supersymmetry transformations. 
{Recall that the massless Rarita-Schwinger equation has a fermionic gauge symmetry.  Namely, it is invariant under the gauge transformation
\begin{equation}\label{LargeSusy}
    \Psi_{\mu} \mapsto \Psi_{\mu}+\nabla_\mu \lambda\,,
\end{equation}
where $\lambda$ is an anti-commuting spinor. This is the local supersymmetry of supergravity.\footnote{{We assume a purely bosonic asymptotically flat background and have dropped the corresponding transformation on the frame field. Note that while the spinor $\lambda$ here is anti-commuting, in the rest of the section we restrict to commuting component spinors which should be dressed with Grassmann variables to compare to~\eqref{LargeSusy}.}} When the asymptotic behavior of $\lambda$ is such that the inhomogeneous shift goes to an arbitrary function in $(z,\bz)$ near null infinity, the gauge transformation~\eqref{LargeSusy} corresponds to a (spontaneously broken) {\it large} supersymmetry transformation and the Goldstino is the (conformally) soft gravitino.}

\subsection{Leading Conformally Soft Gravitino}\label{sec:confsoftgravitinos}

For conformal dimension $\Delta=\frac{1}{2}$ the left-handed conformal primary wavefunction with $J=+\frac{3}{2}$ reduces to pure gauge~\cite{Pasterski:2020pdk,Pasterski:2021fjn}
\begin{equation}\label{chisoft}
    \chi^\Gold_{\frac{1}{2},+\frac{3}{2};\mu}=\nabla_\mu \Lambda_{\frac{1}{2},+\frac{3}{2}}\,, \quad \Lambda_{\frac{1}{2},+\frac{3}{2}}=(\epsilon_+\cdot X)o\varphi^{\frac{1}{2}}\,.
\end{equation} 
Besides the harmonic and radial gauge conditions, $\nabla_\mu \chi_\nu=0$ and $X^\mu \chi_\mu=0$ , the conformally soft gravitino~\eqref{chisoft} satisfies the gauge condition $\sb^\mu\chi_\mu=0$~\cite{Lysov:2015jrs}.
At future null infinity its angular components take the form 
\be\label{chisoftscri}
\chi^\Gold_{\frac{1}{2},+\frac{3}{2};z}
=\frac{-i}{(z-w)^2}\left(
\begin{array}{c}
\bw \\ -1
\end{array}\right)
\,,\quad 
\chi^\Gold_{\frac{1}{2},+\frac{3}{2},\bz}
=2\pi i \delta^{(2)}(z-w)\left(
\begin{array}{c}
\bw \\ -1
\end{array}\right)
\,,
\ee
while its temporal and radial components behave, respectively, as $\O(1/r)$ and $\O(1/r^2)$. Hence the gravitino wavefunction $\chi_{\frac{1}{2},+\frac{3}{2};\mu}$ obeys the standard fall-off conditions and obviously has vanishing `field strength' $\mathcal{f}_{\mu\nu}=\nabla_\mu \chi_\nu-\nabla_\nu \chi_\mu$. Comparing this to~\eqref{LargeSusy} we recognize the $(\Delta,J)=(\frac{1}{2},+\frac{3}{2})$ gravitino~\eqref{chisoftscri} as the Goldstino for spontaneously broken large supersymmetry. 

Related to the conformally soft gravitino~\eqref{chisoft} by a shadow transform is the conformal shadow primary of spin $J=-\frac{3}{2}$ and conformal dimension $\Delta=\frac{3}{2}$ which also reduces to pure gauge, albeit with a more complicated potential,
\begin{equation}\label{tchisoft}
  \widetilde{\chi}^\Gold_{\frac{3}{2},-\frac{3}{2};\mu}=\nabla_\mu \Lambda_{\frac{3}{2},-\frac{3}{2}}\,, \quad \Lambda_{\frac{3}{2},-\frac{3}{2}}=\sqrt{2}\left((\epsilon_-\cdot X)\iota-{\textstyle \frac{1}{2}}(\epsilon_-\cdot X)^2o\right)\varphi^{\frac{3}{2}}\,,
\end{equation}
and satisfies the harmonic and radial gauge condition as well as $\bar{\sigma}^\mu \tchi_\mu=0$.
At future null infinity its angular components take the form\footnote{Using~\eqref{2dShadowTransform} it is straightforward to check that the leading $z$ component in the large-$r$ expansion of~\eqref{tchisoft} is the shadow transform of~\eqref{chisoft}, while to show this for the $\bz$ component it is convenient to write~\eqref{chisoft} as a sum over (derivatives of) conformal integrals and use~\eqref{I2} to solve them.} 
\be\label{tchisoftscri}
\widetilde{\chi}^\Gold_{\frac{3}{2},-\frac{3}{2};z}
=\pi i \delta^{(2)}(z-w)
\left(\begin{array}{c}
1 \\ 0 
\end{array}\right)
-\pi i \partial_\bz \delta^{(2)}(z-w) 
\left(\begin{array}{c}
\bz \\ -1 
\end{array}\right)
\,,\quad
\widetilde{\chi}^\Gold_{\frac{3}{2},-\frac{3}{2};\bz}
=\frac{-i}{(\bz-\bw)^3}
\left(
\begin{array}{c}
\bz \\ -1 
\end{array}\right)
\,,
\ee
while its temporal and radial components behave, respectively, as $\O(1/r)$ and $\O(1/r^2)$ and the `field strength' again vanishes. We recognize the $(\Delta,J)=(\frac{3}{2},-\frac{3}{2})$ gravitino~\eqref{tchisoftscri} as the (shadow) Goldstino for spontaneously broken large supersymmetry.

\subsection{Celestial Gravitino Diamond}
\begin{figure}[ht!]
\centering
\begin{tikzpicture}[scale=1.2]
\definecolor{red}{rgb}{.5, 0.5, .5};
\draw[thick,->] (-1+.05,1-.05)node[left]{${\tchi}^{\Gold}_{\frac{3}{2},-\frac{3}{2}}$} --node[below left]{} (-.05,.05) ;
\draw[thick,->] (2-.05,2-.05) node[right]{${\chi}^{\Gold}_{\frac{1}{2},+\frac{3}{2}}$} --node[below right]{} (1+.1414/2,1+.1414/2);
\draw[thick,->] (1-.1414/2,1-.1414/2)-- (.05,.05);
\filldraw[black] (0,0) circle (2pt) node[below]{$\chi^{gen,\Gold}_{\frac{5}{2},-\frac{1}{2}}~~~$};
\filldraw[white] (0,0) circle (1pt) ;
\filldraw[black] (-1,1) circle (2pt) ;
\filldraw[black] (2,2) circle (2pt) ;
\draw[thick] (1+.1414/2,1+.1414/2) arc (45:-135:.1);
\draw[thick] (0+.1414/2,2+.1414/2) arc (45:-135:.1);
\draw[thick,->] (1,3)  node[above,black]{$~~~\chi^{gen,\Gold}_{-\frac{1}{2},+\frac{1}{2}}$} -- (2-.05,2+.05);
\draw[thick,->] (0-.1414/2,2-.1414/2)-- (-1+.05,1.05);
\draw[thick,->] (1,3) -- (0+.1414/2,2+.1414/2);
\node[fill=black,regular polygon, regular polygon sides=4,inner sep=1.6pt] at (1,3) {};
\node[fill=white,regular polygon, regular polygon sides=4,inner sep=.8pt] at (1,3) {};
\filldraw[red,thick] (1,0) circle (2pt);
\filldraw[white] (1,0) circle (1pt) node[below,black]{$~~~{\bchi}^{gen,\Gold}_{\frac{5}{2},+\frac{1}{2}}$};
\filldraw[red,thick] (-1,2) circle (2pt) node[left,black]{$-{\bchi}^{\Gold}_{\frac{1}{2},-\frac{3}{2}}$};
\draw[red, thick] (0-.1414/2,1+.1414/2) arc (135:315:.1);
\draw[red, thick] (1-.1414/2,2+.1414/2) arc (135:315:.1);
\draw[->,red,thick] (-1,2) --  (0-.1414/2,1+.1414/2);
\draw[->,red,thick] (0+.1414/2,1-.1414/2) -- (1-.05,0.05);
\filldraw[red,thick] (2,1) circle (2pt) ;
\draw[->,red,thick] (2,1)  node[right,black]{${\tbchi}^{\Gold}_{\frac{3}{2},+\frac{3}{2}}$} -- (1+.05,0.05);
\draw[->,red,thick] (0,3) node[above,black]{${\bchi}^{gen,\Gold}_{-\frac{1}{2},-\frac{1}{2}}~~~$}  --  (1-.1414/2,2+.1414/2);
\draw[->,red,thick] (1+.1414/2,2-.1414/2) --  (2-.05,1+.05);
\draw[->,red,thick] (0,3) --  (-1+.05,2+.05);
\node[fill=red,regular polygon, regular polygon sides=4,inner sep=1.6pt] at (0,3) {};
\node[fill=white,regular polygon, regular polygon sides=4,inner sep=.8pt] at (0,3) {};
\end{tikzpicture}
\caption{Goldstino diamond for the leading soft gravitino theorem~\cite{Pasterski:2021fjn}.
}
\label{fig:gravitinodiamond}
\end{figure}
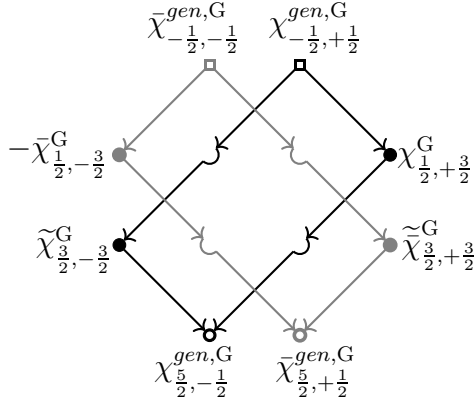

The conformally soft gravitinos~\eqref{chisoft} and~\eqref{tchisoft} form the left and right corners of the celestial gravitino diamond shown in figure~\ref{fig:gravitinodiamond} and descend to the same generalized conformal primary at the bottom of the diamond 
\begin{equation}
  \partial_w  \tchi^\Gold_{\frac{3}{2},-\frac{3}{2}}=\chi^{gen,\Gold}_{\frac{5}{2},-\frac{1}{2}} =\frac{1}{2!} \partial_\bw^2 \chi^\Gold_{\frac{1}{2},+\frac{3}{2}}\,.
\end{equation}
We can complete the diamond at the top corner with a generalized conformal primary wavefunction that descends to the radiative conformal primaries\footnote{The ambiguity in defining the top corner is discussed in~\cite{Pasterski:2021fjn}.}
\begin{equation}
  \tchi^\Gold_{\frac{3}{2},-\frac{3}{2}}=\frac{1}{2!}\partial_\bw^2  \chi^{gen,\Gold}_{-\frac{1}{2},+\frac{1}{2}}\,, \quad \chi^\Gold_{\frac{1}{2},+\frac{3}{2}}=\partial_w  \chi^{gen,\Gold}_{-\frac{1}{2},+\frac{1}{2}}\,.
\end{equation}
This is summarized in table~\ref{table:Ggravitinos}.

\vspace{1em}
\begin{table}[ht!]
\renewcommand*{\arraystretch}{1.3}
\centering
    \begin{tabular}{l|l|l|l|l}
     Corner & $\Delta$  & $J$ 
       &  $\chi^{\Gold}_{\Delta,J}$ & $\Lambda_{\Delta,J}$\\
         \hline
         Top&$-\frac{1}{2}$&$+\frac{1}{2}$
         & $\frac{1}{\sqrt{2}}  ol_\mu\varphi^{-\frac{1}{2}}$&$- \frac{1}{\sqrt{2}}o\varphi^{-\frac{1}{2}}\log\varphi^{-1}$\\
          Left&$~~\frac{3}{2}$&$-\frac{3}{2}$
          &$\sqrt{2}\iota{\bar m}_\mu \varphi^{\frac{3}{2}}$&$\frac{1}{2!}\p_\bw^2\Lambda_{-\frac{1}{2},+\frac{1}{2}}$\\
           Right&$~~\frac{1}{2}$&$+\frac{3}{2}$
           &$o m_\mu \varphi^\frac{1}{2}$&$\p_w\Lambda_{-\frac{1}{2},+\frac{1}{2}}$\\
           Bottom&$~~\frac{5}{2}$&$-\frac{1}{2}$
           &${2}\left[\left(\frac{X^2}{2}l_\mu+n_\mu\right) \iota +\frac{X^2}{2} o\bar{m}_\mu\right] \varphi^{\frac{5}{2}}$\,
           &$\frac{1}{2!}\p_w\p_\bw^2\Lambda_{-\frac{1}{2},+\frac{1}{2}}$\\
    \end{tabular}
    \caption{Elements of the celestial diamond corresponding to large supersymmetry~\cite{Pasterski:2020pdk,Pasterski:2021fjn}.}
    \label{table:Ggravitinos}
\end{table}

\subsection{Soft Charge for Large Supersymmetry}\label{sec:gravitinocharge}

In the language of the covariant phase space formalism, computing the soft part of the large supersymmetry charge amounts to computing the symplectic structure for a generic gravitino perturbation with radiative fall-offs at null infinity and the Goldstino associated to large supersymmetry. This corresponds to the 2D operator~\eqref{Ffermionic} for~\eqref{Fgravitinos} with the conformally soft gravitinos~\eqref{chisoftscri} or~\eqref{tchisoftscri}.\footnote{See appendix~\ref{app:Omega} for details.  Comparing to the results of appendix~\ref{app:IP}, we see that  \be\mathcal{O}^{s,\pm}_{\Delta,J}(w,\bw)\equiv i(\hat{\Psi}^s(X^\mu),\Psi^s_{\Delta^*,-J}(X_\mp^\mu;w,\bw))=\Omega(\hat{\Psi}^s(X^\mu),\Psi^s_{\Delta,J}(X_\pm^\mu;w,\bw))\,,\ee
 after an appropriate complexification of the symplectic product to allow pairings between the Majorana field operators and the primaries~\eqref{chiralspinor}.}
The final result for the soft charge of spontaneously broken large supersymmetry is given by (expression~\eqref{Omegaleadingchi} evaluated at null infinity).
For the left handed Goldstino diamond, only the first term in~\eqref{Fgravitinos} contributes and we have (omitting the $s$ label)
\begin{equation}\label{softgravitinocharge}
   \O_{\Delta,J}=i\int du d^2z \,\hat{\chi}^{\dagger(0)}_z \sb_u \chi^\Gold_{\Delta,J;\bz}\,,
\end{equation}
generating the shift~\eqref{shift2}. This yields $\O_{\frac{1}{2},+\frac{3}{2}}$ for the Goldstino
\begin{equation}\label{chicontact}
   \chi^\Gold_{\frac{1}{2},+\frac{3}{2},\bz}
=2\pi i \delta^{(2)}(z-w)\left(
\begin{array}{c}
\bw \\ -1
\end{array}\right)\,,
\end{equation}
which establishes the equivalence between the large supersymmetry Ward identity and the (conformally) soft gravitino theorem~\cite{Fotopoulos:2020bqj}. We see this as follows.  After using the inverse Mellin transform to express the creation and annihilation operators in~\eqref{chiScri} in terms of $a_{\Delta,+\frac{3}{2}}(z,\bz)$ and $a^\dagger_{\Delta,+\frac{3}{2}}(z,\bz)$, the $\omega$-integral takes the form
\begin{equation}
    \int_0^\infty d\omega \omega^{\frac{1}{2}-\Delta} e^{\pm i \omega(1+z\bz)u_\pm}=(\mp i)^{\Delta-\frac{3}{2}}{\textstyle \Gamma(\frac{3}{2}-\Delta)} u_\pm^{\Delta-\frac{3}{2}}(1+z\bz)^{\Delta-\frac{3}{2}}\,,
\end{equation}
where we analytically continue $u\mapsto u_\pm =u\pm i\varepsilon$ to guarantee convergence.  Then using the generalized distribution~\cite{Donnay:2020guq}
\begin{equation}
    \int_0^\infty du u^{\Delta-\frac{3}{2}}=2\pi {\textstyle \boldsymbol{\delta}(i(\Delta-\frac{1}{2}))}\,,
\end{equation}  we find that the $u$-integrals in the soft charge ${\cal O}_{\Delta,J}$ for $\Delta=\frac{1}{2}$, $J=+\frac{3}{2}$ gives\footnote{Setting $\Delta=\frac{1}{2}$ before evaluating the integral in~\eqref{lightray} gives the same factor of $\frac{1}{2}$ that was encountered in the Ward identity papers~\cite{He:2014laa}.   This is effectively taking the average of the $u_+$ and $u_-$ integrals, giving us both positive and negative frequency contributions for the gravitino zero modes.}
\begin{equation}
    \int_0^\infty d\omega \omega^{\frac{1}{2}} a_+(\omega)\int_{-\infty}^{+\infty} du e^{-i \omega(1+z\bz)u_-}=\pi \lim_{\Delta \to \frac{1}{2}} {\textstyle (\Delta -\frac{1}{2})} a_{\Delta,+\frac{3}{2}} (1+z\bz)^{-1}\,,
\end{equation}
and
\begin{equation}
    \int_0^\infty d\omega \omega^{\frac{1}{2}} a^\dagger_-(\omega)\int_{-\infty}^{+\infty} du e^{+i \omega(1+z\bz)u_+}=\pi \lim_{\Delta \to \frac{1}{2}} {\textstyle (\Delta -\frac{1}{2})} a^\dagger_{\Delta,+\frac{3}{2}} (1+z\bz)^{-1}\,.
\end{equation}
 Plugging this into~\eqref{softgravitinocharge} and integrating over $(z,\bz)$ yields the soft charge 
 \begin{equation}\label{contact}
    \O_{\frac{1}{2},+\frac{3}{2}}(w,\bw)={\textstyle\frac{i\kappa}{2}}\lim_{\Delta \to \frac{1}{2}} {\textstyle \left(\Delta-\frac{1}{2}\right)}( a_{\Delta,+\frac{3}{2}}-a^\dagger_{\Delta,+\frac{3}{2}})\,.
\end{equation}
Meanwhile, the 2D operator $\tO_{\frac{3}{2},-\frac{3}{2}}(w,\bw)$ obtained from~\eqref{softgravitinocharge} for the $\Delta=\frac{3}{2}$ shadow Goldstino 
\begin{equation}\label{tchinoncontact}
 \widetilde{\chi}^\Gold_{\frac{3}{2},-\frac{3}{2};\bz}
=\frac{-i}{(\bz-\bw)^3}
\left(
\begin{array}{c}
\bz \\ -1 
\end{array}\right)\,,
\end{equation}
gives rise to the $(h,\bh)=(0,\frac{3}{2})$ supercurrent.

This is similar to the situation in gravity where\footnote{The Bondi news is $N_{AB}=\lim\limits_{r\to \infty}\frac{1}{r}\p_u h_{AB}$.} the $\Delta=0$ Goldstone mode $N^\Gold_{0,+2;\bz\bz}=-2\pi\delta^{(2)}(z-w)$ generating Diff($S^2$) symmetry yields equivalence with the subleading (conformally) soft graviton theorem while the $\Delta=2$ shadow Goldstone mode $\tN^\Gold_{2,-2;\bz\bz}=-\frac{1}{(\bz-\bw)^4}$ gives rise to the $(h,\bh)=(0,2)$ stress tensor~\cite{Donnay:2020guq}. The soft charges generated by these superrotation and Diff($S^2$) modes were shown to be related by a 2D shadow transform in celestial CFT. Here, similarly, the soft charges for the Goldstinos~\eqref{chicontact} and~\eqref{tchinoncontact} are related by the shadow transform~\eqref{2dShadowTransform} as 
\begin{equation}
\widetilde{\O}_{\frac{3}{2},-\frac{3}{2}}(w,\bw)={\frac{1}{2\pi}}\int d^2w' \frac{1}{(\bw-\bw')^3} \O_{\frac{1}{2},+\frac{3}{2}}(w',\bw')\,,
\end{equation}
and the inverse
\begin{equation}
\O_{\frac{1}{2},+\frac{3}{2}}(w,\bw)={-}\frac{1}{\pi}\int d^2w' \frac{(\bw-\bw')}{(w-w')^2} \widetilde{\O}_{\frac{3}{2},-\frac{3}{2}}(w',\bw')\,.
\end{equation}

\subsection{Soft Operator in the Gravitino Diamond}\label{sec:softOpgravitino}

Starting from~\eqref{contact}, we see much like the superrotation/Diff$(S^2)$ case that one of the corners of the diamond is isomorphic to the soft theorem.  We can use descendancy relations to take us to the bottom corner of the diamond. Letting 
\be\label{osoft}
[\O_{soft}|_{\bz}^{a}= -\frac{\pi}{\sqrt{2}} D_{\bz}^2 \int du \hat{\chi}^\dagger_{z\dot{c}} \bar{\sigma}^{\dot{c} a}_u\,,
\ee
we construct the object
\begin{equation}\label{feta}
    \Q(\eta)=\int d^2z \,[\O_{soft}|\eta]\,,
\end{equation}
where we suppress the tensor index contractions between the soft operator~\eqref{osoft} and spinor $\eta^{\bz}_{a}$ since $\eta^{z}_{a}=0$. In terms of the creation and annihilation operators 
 \begin{equation}\label{Osoft}
 \badat{3}
 [\O_{soft}|&={\textstyle  \frac{i \kappa}{8}}\lim_{\Delta \to \frac{1}{2}} {\textstyle \left(\Delta-\frac{1}{2}\right)}D_\bz^2\left[(1+z\bz)^{-1}( a_{\Delta,+\frac{3}{2}}-a^\dagger_{\Delta,+\frac{3}{2}})\langle x|\bar{\sigma}_u\right]\,,\\
 &={\textstyle \frac{i \kappa}{8}}\lim_{\Delta \to \frac{1}{2}} {\textstyle \left(\Delta-\frac{1}{2}\right)}(1+z\bz)^{-1}\partial_\bz^2\left[( a_{\Delta,+\frac{3}{2}}-a^\dagger_{\Delta,+\frac{3}{2}})\right]\langle x|\bar{\sigma}_u\,.
 \eadat
 \end{equation}
We see that the operator $[\mathcal{O}_{soft}|x]$ is a level-2 primary descendant operator with $\Delta=\frac{5}{2}$ and $J=-\frac{1}{2}$, and the smeared operator~\eqref{feta} gives the fermionic analogue of the charge operators of~\cite{Banerjee:2018fgd,Banerjee:2019aoy,Banerjee:2019tam}. Much like the spin-1 and spin-2 cases studied in~\cite{Pasterski:2021dqe} the spacetime descendants on the round sphere map to flattened celestial sphere descendants of the Mellin transformed modes. 
 
Moreover, we see from~\eqref{Osoft} that 
 \be
 {\cal Q}(\frac{\bw-\bz}{w-z}|x])=
 \pi{\cal O}_{\frac{1}{2},+\frac{3}{2}},~~~
 {\cal Q}(\frac{1}{\bw-\bz}|x])=2\pi\widetilde{\cal O}_{\frac{3}{2},-\frac{3}{2}}\,,
 \ee
while
   \be
 {\cal Q}(\delta^{(2)}(z-w)|x])=\p_w{{\widetilde{\cal O}}}_{\frac{3}{2},-\frac{3}{2}}=\frac{1}{2!}\p_\bw^2{\cal O}_{\frac{1}{2},+\frac{3}{2}}.
 \ee
This form of the charge where $\eta=\varepsilon(z,\bz)|x]$ is consistent with that of~\cite{Avery:2015gxa,Lysov:2015jrs}. We are able to use $\eta$ with such a simple form because there are various kernels of the integrated charge~\eqref{feta} coming from the kernels of the descendancy relations of the celestial diamond, as well as that of the spinor product.\footnote{Note that the gauge potential of the conformally soft gravitino~\eqref{chisoft} at null infinity can indeed be written as
\begin{equation}\label{eta}
    \Lambda_{\frac{1}{2},+\frac{3}{2}}|_{\mathcal{I}^+}\simeq D_A \eta^A_+\,,\quad
    \eta^z_+=0\,, \quad \eta^\bz_+=i\frac{\bz-\bw}{z-w}\frac{1+z\bz}{1+w\bw}\left(\begin{array}{c} \bw\\-1\end{array}\right)\,,
\end{equation}
where the equivalence is up to the kernel of the sphere derivatives giving $\chi_{A}$. The angular components of the Goldstino can then be expressed as
\begin{equation}
    \chi^G_{\frac{1}{2},+\frac{3}{2};z}=D_z D_\bz \eta^\bz_+\,, \quad \chi^G_{\frac{1}{2},+\frac{3}{2};\bz}=D_\bz^2 \eta^\bz_+\,.
\end{equation}
The soft charge~\eqref{softgravitinocharge} takes the form
\begin{equation}
    \O_{\frac{1}{2},+\frac{3}{2}}={i}\int du d^2z  \, D_\bz^2\,\hat{{\chi}}^{\dagger(0)}_{z}\, \bar \sigma_u \eta^\bz_+ \,.
\end{equation}}

 The soft operator lies at the bottom of the subleading soft graviton memory diamonds, conveniently captured by the following diagram.
\begin{equation}
  \raisebox{-2cm}{
\begin{tikzpicture}[scale=0.6]
\definecolor{red}{rgb}{.5, 0.5, .5};
\draw[thick,->] (-1+.05,1-.05)node[left]{$\tO_{\frac{3}{2},-\frac{3}{2}}=\frac{1}{2!}\p_\bw^2 \O_{-\frac{1}{2},+\frac{1}{2}}$} --node[below left]{} (-.05,.05) ;
\draw[thick,->] (2-.05,2-.05) node[right]{$\O_{\frac{1}{2},+\frac{3}{2}}=\p_w \O_{-\frac{1}{2},+\frac{1}{2}}$} --node[below right]{} (1+.1414/2,1+.1414/2);
\draw[thick,->] (1-.1414/2,1-.1414/2)-- (.05,.05);
\filldraw[black] (0,0) circle (2pt) node[below]{$ [\O_{soft}|x]_{\frac{5}{2},-\frac{1}{2}}~~~$}; 
\filldraw[white] (0,0) circle (1pt) ;
\filldraw[black] (-1,1) circle (2pt) ;
\filldraw[black] (2,2) circle (2pt) ;
\draw[thick] (1+.1414/2,1+.1414/2) arc (45:-135:.1);
\draw[thick] (0+.1414/2,2+.1414/2) arc (45:-135:.1);
\draw[thick,->] (1,3)  node[above,black]{$~~~\O_{-\frac{1}{2},+\frac{1}{2}}$} -- (2-.05,2+.05);
\draw[thick,->] (0-.1414/2,2-.1414/2)-- (-1+.05,1.05);
\draw[thick,->] (1,3) -- (0+.1414/2,2+.1414/2);
\node[fill=black,regular polygon, regular polygon sides=4,inner sep=1.6pt] at (1,3) {};
\node[fill=white,regular polygon, regular polygon sides=4,inner sep=.8pt] at (1,3) {};
\end{tikzpicture}
}
\end{equation}
As shown in the diagram, from the definitions of the large supersymmetry current and its shadow we can formally write $\tO_{\frac{3}{2},-\frac{3}{2}}$ and $\O_{\frac{1}{2},+\frac{3}{2}}$ as descendants of $\O_{-\frac{1}{2},+\frac{1}{2}}$, a generalized primary operator with conformal dimension $\Delta=-\frac{1}{2}$ and spin $J=+\frac{1}{2}$ that lies at the top of the gravitino memory diamond.  This operator is formally defined via
\begin{equation}
    \O_{-\frac{1}{2},+\frac{1}{2}}=i(\hat{\Psi}^s,\Psi^{s=\frac{3}{2}}_{\frac{1}{2},-\frac{1}{2}})\,,~~~~~~\Psi^{s=\frac{3}{2}}_{\frac{1}{2},-\frac{1}{2}}=\left(\begin{array}{c}
    0  \\
    \frac{1}{\sqrt{2}}\bar{o}l_\mu\varphi^{-\frac{1}{2}} 
    \end{array}\right).
\end{equation}
This operator is strictly speaking not part of the spectrum but its role is important for the theory, e.g. it can be used to define the other primary operators and (primary) descendants in the theory -- see~\cite{Pasterski:2021fjn,Pasterski:2021dqe} for a related discussion. 

We close this section by noting that the 
the spin~1,$\frac{3}{2}$ and~2 symmetry currents 
\be\label{symcurr}
\widetilde{\cal O}_{1,-1},\widetilde{\cal O}_{\frac{3}{2},-\frac{3}{2}},\widetilde{\cal O}_{2,-2}\,,
\ee
corresponding to large U(1), large SUSY and superrotation symmetry, respectively, all have the same symmetry parameter  $\propto \frac{1}{\bz-\bw}$ (called $\varepsilon$ and $Y^z$ for spin~1 and~2 in~\cite{Pasterski:2021dqe} and $\eta$ for spin~$\frac{3}{2}$ above), for their appropriately normalized charges, since their first $\p_w$ descendants are the respective soft charges. We will investigate the relationship between the descendancy relations in celestial diamonds and the spin-shifting relations of SUSY in section~\ref{celestialpyramids}.

\section{Soft Charges without Goldstinos}\label{sec:subgoldstinos}

In this section we examine the degenerate celestial diamonds corresponding to the soft photino\footnote{Generalizing to the gluino case is straightforward for the soft charges constructed herein. Since they are linear in the perturbation, this amounts to adding a Lie algebra index.} and subleading soft gravitino. 

\subsection{Conformally Soft Photino}\label{sec:photino}
\begin{figure}[h!]
\centering
\begin{tikzpicture}[scale=.9]
\filldraw[black] (1,1) circle (2pt) ;
\filldraw[black] (0,2) circle (2pt) ;
\filldraw[black] (0,1) circle (2pt) ;
\filldraw[black] (1,2) circle (2pt) ;
\draw[thick,->] (1,2)-- (0+.05,1+.05);
\draw[thick,->] (0,2)-- (1-.05,1+.05);
\filldraw[black] (0,2) circle (2pt) ;
\node at (-1,1) {$\frac{3}{2}$};
\node at (-1,2) {$\frac{1}{2}$};
\node at (-0.6,3) {$J$};
\node at (-1,2.87) {$\tiny{\Delta}$};
\draw[thick] (-1+.01,3+.2) -- (-1+.28,3-.3);
\node at (0,3) {$-\frac{1}{2}~$};
\node at (1,3) {$\frac{1}{2}$};
\end{tikzpicture}
    \caption{Photino `diamonds'.}
\label{fig:photinodiamond}
\end{figure}
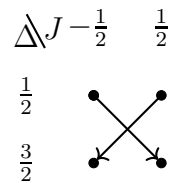

The conformally soft spin $s=\frac{1}{2}$ Weyl spinor $\psi_{\frac{1}{2},+\frac{1}{2}}$ and its shadow $\tpsi_{\frac{3}{2},-\frac{1}{2}}$ correspond to the zero-area celestial diamonds shown in figure~\ref{fig:photinodiamond}.
Using appendix~\ref{app:Omega} we can write a 2D operator analogous to the soft charge~\eqref{softgravitinocharge}. Namely, evaluating expression~\eqref{Omegapsi} at null infinity yields
\begin{equation}\label{SoftChargePhotino}
    \O_{\frac{1}{2},+\frac{1}{2}}=i\int_{\I^+} du d^2z \sqrt{\gamma}
 \left[\hat{\psi}^{\dagger(1)} \sb_u \psi_{\frac{1}{2},+\frac{1}{2}}^{(1)}\right]\,,
\end{equation}
where the superscript on the photinos indicates the coefficient of the $r^{-1}$ term in their large-$r$ expansions. The form of~\eqref{SoftChargePhotino} is consistent with the soft charge in~\cite{Dumitrescu:2015fej}.
At coincident points on the celestial sphere the contribution from the conformally soft photino is
\begin{equation}\label{confsoftpsiScricontact}
    \psi_{\frac{1}{2},+\frac{1}{2}}^{(1)}\Big|_{z=w}=i\pi\lim_{\Delta\to\frac{1}{2}}\frac{u^{\frac{1}{2}-\Delta}}{\Delta-\frac{1}{2}}(1+z\bz)\delta^{(2)}(z-w)|x]\,,
\end{equation}
while at non-coincident we have
\begin{equation}\label{confsoftpsiScrinoncontact}
 \psi_{\frac{1}{2},+\frac{1}{2}}^{(1)}\Big|_{z\neq w}=\frac{i}{2}\frac{1+z\zb}{(z-w)(\zb-\bw)}|q]\,.
\end{equation}
Analogous expressions exist for the opposite helicity photino yielding $\O_{\frac{1}{2},- \frac{1}{2}}$. Furthermore, as indicated in figure~\eqref{fig:photinodiamond} the $\Delta=\frac{1}{2}$ conformally soft photino descend to the $\Delta=\frac{3}{2}$ shadow photino as
\begin{equation}
    \p_\bw \psi_{\frac{1}{2},+\frac{1}{2}}=-\tpsi_{\frac{3}{2},-\frac{1}{2}}\,,
    \quad     \p_w \bar{\psi}_{\frac{1}{2},-\frac{1}{2}}=-\widetilde{\bar{\psi}}_{\frac{3}{2},+\frac{1}{2}}\,,
\end{equation}
for which we can write the 2D operators $\tO_{\frac{3}{2},\pm \frac{1}{2}}$.

Note that the isomorphism with the conformally soft photino theorem~\cite{Fotopoulos:2020bqj} follows from the contact term~\eqref{confsoftpsiScricontact}. Indeed, if we renormalize the operator by $(\Delta-\frac{1}{2})$, the soft charge becomes\footnote{Without renormalizing the limit $\Delta \to \frac{1}{2} $ furthermore yields a contribution to~\eqref{confsoftpsiScricontact} involving a logarithm in~$u$ which will be discussed elsewhere.}
\begin{equation}
    \O^{ren}_{\frac{1}{2},+\frac{1}{2}}(w,\bw)
    =- 2 \pi(1+w\bw)^{-1}  \int du 
    \hat{\psi}^{\dagger(1)}\sb_u |q]
 ={\frac{ie}{2} }\lim_{\Delta \to \frac{1}{2}} (\Delta-\frac{1}{2}) (a_{\Delta,+\frac{1}{2}}-a^\dagger_{\Delta,+\frac{1}{2}})\,.
\end{equation}

\subsection{Subleading Conformally Soft Gravitino}\label{sec:subgravitino}
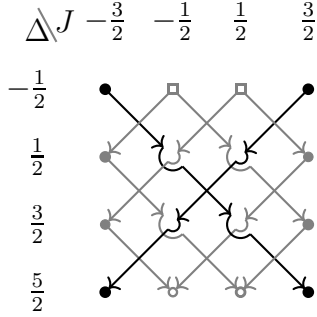
\begin{figure}[h!]
\centering
\begin{tikzpicture}[scale=.9]
\definecolor{red}{rgb}{.5, 0.5, .5};
\draw[thick,red,->] (-1+.05,1-.05)node[left]{
} --node[below left]{
} (-.05,.05) ;
\draw[thick,red,->] (2-.05,2-.05)node[above]{
} --node[below right]{
} (1+.1414/2,1+.1414/2);
\draw[thick,red,->] (1-.1414/2,1-.1414/2)-- (.05,.05);
\filldraw[red] (0,0) circle (2pt);
\filldraw[red] (-1,1) circle (2pt) ;
\filldraw[red] (2,2) circle (2pt) ;
\draw[red,thick] (1+.1414/2,1+.1414/2) arc (45:-135:.1);
\draw[red,thick] (0+.1414/2,2+.1414/2) arc (45:-135:.1);
\node at (-2,0) {$\frac{5}{2}$};
\node at (-2,1) {$\frac{3}{2}$};
\node at (-2,2) {$\frac{1}{2}$};
\node at (-2,3) {$-\frac{1}{2}~~$};
\node at (-1.6,4) {$J$};
\node at (-2,3.87) {$\tiny{\Delta}$};
\draw[red,thick] (-2+.01,4+.2) -- (-2+.28,4-.3);
\node at (-1,4) {$-\frac{3}{2}$};
\node at (0,4) {$-\frac{1}{2}$};
\node at (1,4) {$\frac{1}{2}$};
\node at (2,4) {$\frac{3}{2}$};
\filldraw[black] (1,0) circle (2pt) ;
\filldraw[red,thick] (1,0) circle (2pt) ;
\filldraw[red,thick] (-1,2) circle (2pt) ;
\draw[red, thick] (0-.1414,1+.1414) arc (135:315:.2);
\draw[red, thick] (1-.1414,2+.1414) arc (135:315:.2);
\draw[black, thick] (0-.1414,2+.1414) arc (135:315:.2);
\draw[black, thick] (1-.1414,1+.1414) arc (135:315:.2);
\draw[->,red,thick] (-1,2) --  (0-.1414,1+.1414);
\draw[->,red,thick] (0+.1414,1-.1414) -- (1-.05,0.05);
\filldraw[red,thick] (2,1) circle (2pt) ;
\filldraw[black,thick] (-1,0) circle (2pt) ;
\filldraw[black,thick] (-1,3) circle (2pt) ;
\filldraw[black,thick] (2,3) circle (2pt) ;
\filldraw[black,thick] (2,0) circle (2pt) ;
\draw[->,thick] (-1,3) -- (0-.1414,2+.1414);
\draw[->,thick] (1+.1414,1-.1414) -- (2-.05,0+.05);
\draw[->,thick] (0+.1414,2-.1414) -- (1-.1414,1+.1414);
\draw[black,thick] (0+.1414/2,1+.1414/2) arc (45:-135:.1);
\draw[black,thick] (1+.1414/2,2+.1414/2) arc (45:-135:.1);
\draw[->,thick] (0-.1414/2,1-.1414/2) -- (-1+.05,0+.05);
\draw[->,thick] (1-.06,2-.06) -- (0+.06,1+.06);
\draw[->,thick] (2,3) -- (1+.06,2+.06);
\draw[->,red,thick] (2,1) -- (1+.05,0.05);
\draw[->,red,thick] (0,3) --  (1-.1414,2+.1414);
\draw[->,red,thick] (1+.1414,2-.1414) --  (2-.05,1+.05);
\draw[->,red,thick] (0,3) --  (-1+.05,2+.05);
\draw[red,thick,->] (1,3)-- (2-.05,2+.05);
\filldraw[black] (1,3) circle (2pt) ;
\draw[red,thick,->] (0-.1414/2,2-.1414/2)-- (-1+.05,1.05);
\draw[red,thick,->] (1,3) -- (0+.1414/2,2+.1414/2);
\filldraw[red,thick] (0,3) circle (2pt) ;
\filldraw[white] (1,0) circle (1pt) ;
\filldraw[white] (0,0) circle (1pt) ;
\node[fill=red,regular polygon, regular polygon sides=4,inner sep=1.6pt] at (0,3) {};
\node[fill=white,regular polygon, regular polygon sides=4,inner sep=.8pt] at (0,3) {};
\node[fill=red,regular polygon, regular polygon sides=4,inner sep=1.6pt] at (1,3) {};
\node[fill=white,regular polygon, regular polygon sides=4,inner sep=.8pt] at (1,3) {};
\end{tikzpicture}
    \caption{Subleading gravitino `diamonds' (black) on top of leading gravitino diamonds (grey).}
\label{fig:subleadinggravitinodiamond}
\end{figure}

The conformally soft spin $s=\frac{3}{2}$ Weyl spinor $\chi_{-\frac{1}{2},+\frac{3}{2};\mu}$ and its shadow $\tchi_{\frac{5}{2},-\frac{3}{2};\mu}$ corresponds to the zero-area celestial diamond shown in figure~\ref{fig:subleadinggravitinodiamond}.
Using again appendix~\ref{app:Omega}, we can write the 2D~operator corresponding to a soft charge for the subleading conformally soft gravitino, namely expression~\eqref{Omegasubleadingchi} evaluated at null infinity. For the gravitino at $\Delta=-\frac{1}{2}$ this yields
\be
\begin{aligned}
\mathcal{O}_{-\frac{1}{2},+\frac{3}{2}}=i\int_{\scri^+} du d^2z &\Big\{\sqrt{\gamma}\left[\delta\chi^{\dagger\, (2)}_r\left(\frac{1}{2}\sb_r-\sb_u\right)\chi^{(0)}_{-\frac{1}{2},+\frac{3}{2};\,u}+\delta\chi^{\dagger\, (1)}_u\left(\frac{1}{2}\sb_r-\sb_u\right)\chi^{(1)}_{-\frac{1}{2},+\frac{3}{2};\,r}\right]\\
&\qquad -\delta\chi^{\dagger\, (1)}_\bz \left(\frac{1}{2}\sb_r-\sb_u\right)\chi^{(-1)}_{-\frac{1}{2},+\frac{3}{2};\,z}+\delta\chi_z^{\dagger\, (0)}\sb_u\chi^{(0)}_{-\frac{1}{2},+\frac{3}{2};\,\bz} \Big\}\,.
\end{aligned}
\ee 
Curiously, while the presence of a soft theorem at this dimension has been noticed in the amplitudes literature~\cite{Liu:2014vva,Fotopoulos:2020bqj}, on the asymptotic symmetry side so far only the large supersymmetry charge for the leading soft gravitino has been computed~\cite{Fuentealba:2021xhn,Avery:2015iix,Lysov:2015jrs}.  
We are happy to contribute a spacetime interpretation of the subleading (conformally) soft gravitino theorem and its corresponding charge. 

At coincident points on the celestial sphere the subleading conformally soft gravitino contributes
\be
\chi_{-\frac{1}{2},+\frac{3}{2};\bz}^{(0)}\Big|_{z=w}=\sqrt{2}\pi i\lim_{\Delta\to-\frac{1}{2}}\frac{u^{\frac{1}{2}-\Delta}}{\Delta+\frac{1}{2}}(1+z\bz)\delta^{(2)}(z-w)|q]\,,
\ee
while all other components are subleading in the large~$r$ expansion. Again the superscript $(n)$ indicates the coefficient of the $r^{-n}$ term. At non-coincident points the conformally soft gravitino has non-radiative fall-offs
\be
\chi_{-\frac{1}{2},+\frac{3}{2};u}^{(0)}\Big|_{z\neq w}=-i\frac{1+\bw z}{\sqrt{2}(z-w)}|q]\,,\quad
\chi_{-\frac{1}{2},+\frac{3}{2};r}^{(1)}\Big|_{z\neq w}=iu\frac{1+\bw z}{\sqrt{2}(z-w)}|q]\,,
\ee
and
\begin{equation}   \chi_{-\frac{1}{2},+\frac{3}{2};z}^{(-1)}\Big|_{z\neq w}=-\sqrt{2}i\frac{\bz-\bw}{(z-w)(1+z \zb)}|q]\,,
\end{equation}
in addition to
 \begin{equation}\scalemath{0.95}{
\chi_{-\frac{1}{2},+\frac{3}{2};z}^{(0)}\Big|_{z\neq w}=iu\frac{(1+\bw z)(1+\bz w)}{(z-w)^2(1+z\zb)}|q]
    \,,\quad   \chi_{-\frac{1}{2},+\frac{3}{2};\bz}^{(0)}\Big|_{z\neq w}=iu\frac{(1+\bw z)^2}{\sqrt{2}(z-w)(\bz-\bw)(1+z\zb)}|q]\,}.
 \end{equation}
Analogous expressions exist for the opposite helicity gravitino yielding $\O_{-\frac{1}{2},- \frac{3}{2}}$. Furthermore, as indicated in figure~\eqref{fig:subleadinggravitinodiamond} the $\Delta=-\frac{1}{2}$ conformally soft gravitinos descend to the $\Delta=\frac{5}{2}$ shadow gravitinos as 
\begin{equation}
    \frac{1}{3!}\p_\bw^3 \chi_{-\frac{1}{2},+\frac{3}{2}}=-\tchi_{\frac{5}{2},-\frac{3}{2}}\,,\quad     \frac{1}{3!}\p_w^3 \bar \chi_{-\frac{1}{2},-\frac{3}{2}}=-\widetilde{\bar{\chi}}_{\frac{5}{2},+\frac{3}{2}}\,,
\end{equation}
for which we can write the 2D operators $\tO_{\frac{5}{2},\pm \frac{3}{2}}$.

As in the case of the photino, the isomorphism with the subleading conformally soft gravitino theorem~\cite{Fotopoulos:2020bqj} follows from the contact term. We can renormalize the charge operator by $(\Delta+\frac{1}{2})$ such that the limit $\Delta \to -\frac{1}{2}$ is finite\footnote{Again, without renormalizing the limit $\Delta \to -\frac{1}{2} $ yields a logarithm in~$u$.}
\begin{equation}
    \O^{ren}_{-\frac{1}{2},+\frac{3}{2}}(w,\bw)=-\sqrt{2}\pi(1+w\wb)\int du u 
\hat\chi_w^{\dagger\, (0)}\sb_u|q]={\frac{\kappa}{2}} \lim_{\Delta \to -\frac{1}{2}}{(\Delta+\frac{1}{2})} (a_{\Delta,+\frac{3}{2}}+a^\dagger_{\Delta,+\frac{3}{2}})\,.
\end{equation}

\section{Celestial Pyramids}\label{celestialpyramids}
In this paper we have completed the soft charge analysis for each of the fermionic celestial diamonds.  This gives us a nice opportunity to merge the investigations of spin-shifting relations in~\cite{Pasterski:2020pdk} and SL($2,\mathbb{C}$) descendants in~\cite{Pasterski:2021dqe}. Namely, when one adds supersymmetry to the mix the celestial diamonds stack into a celestial pyramid.

\begin{figure}[tb!]
    \centering
 \begin{tikzpicture}
 \node (A) at (4,2) {$~~~~~~\mathcal{O}_{-1,2}$};
\node (B) at (3,1) {$~~~~~~\mathcal{O}_{-\frac{1}{2},\frac{3}{2}}$};
\node (C) at (2,0) {$~~~~\mathcal{O}_{0,1}$};
\node (D) at (1, -1) {$~~~~\mathcal{O}_{\frac{1}{2},\frac{1}{2}}$};
\node (E) at (0,-2) {$~~~~\mathcal{O}_{1,0}$};
\node (F) at (4,0) {$~~~~\mathcal{O}_{0,2}$};
\node (G) at (3,-1) {$~~~~\mathcal{O}_{\frac{1}{2},\frac{3}{2}}$};
\node (H) at (4,-2) {$~~~~\mathcal{O}_{1,2}$};
\node (I) at (2,-2) {$~~~~\mathcal{O}_{1,1}$};
 \draw[thick,red,->] (4-.1,2-.1) -- node[above left]{$\overline{Q}$} (3+.2,1+.2);
  \draw[thick,red,->] (3-.1,1-.1) --(2+.2,0+.2);
    \draw[thick,red,->] (2-.1,0-.1) --(1+.2,-1+.2);
     \draw[thick,red,->] (1-.1,-1-.1) --(0+.2,-2+.2);
       \draw[thick,red,->] (4-.1,0-.1) --(3+.2,-1+.2);
       \draw[thick,red,->] (3-.1,-1-.1) --(2+.2,-2+.2);
     \draw[thick, blue, ->] (3+.2,1-.2) -- node[below left]{$Q$} (4-.15,0+.15);
      \draw[thick, blue, ->] (2+.2,0-.2) -- (3-.15,-1+.15);
      \draw[thick, blue, ->] (1+.2,-1-.2) -- (2-.15,-2+.15);
       \draw[thick, blue, ->] (3+.2,-1-.2) -- (4-.15,-2+.15);
     \draw[thick,->] (-1,2) -- node[left]{$\Delta$} (-1,1) ;
     \draw[thick,->] (-1,2) -- node[above]{$J$} (0,2) ;
 \end{tikzpicture}
  \begin{tikzpicture}[scale=.5]
  \node (A) at (-6.5,0){};
    \draw[fill=white!95!gray] (2,2) -- (-4,-4) --  (-2,-6) -- (4,0) --(2,2);
          \draw[fill=white!90!gray] (1,1) -- (-3,-3)-- (-1,-5)--(3,-1) --(1,1);
  
    \draw[fill=white!80!gray] (0,0) -- (-2,-2)-- (0,-4)--(2,-2) --(0,0);

\draw[] (-4,2) --(4,-6);
\draw[] (4,2) --(-4,-6);
 \draw[] (-2,2) -- (4,-4) --  (2,-6) -- (-4,0) --(-2,2);
  \draw[thick] (2,2) -- (-4,-4) --  (-2,-6) -- (4,0) --(2,2);
  \draw[] (0,2) -- (4,-2) --  (0,-6) -- (-4,-2) --(0,2);
  \draw[thick] (1,1) -- (3,-1);
    \draw[thick] (0,0) -- (2,-2);
     \draw[thick] (-2,-2) -- (0,-4);
    \draw[thick] (-3,-3) -- (-1,-5);
  \draw[red, thick ,->] (4,2) -- (3,1);
    \draw[red, thick ,->] (3,1) -- (2,0);
    \draw[red, thick ,->] (2,0) -- (1,-1);
    \draw[red, thick ,->] (1,-1) -- (0,-2);
     \draw[red, thick ,->] (4,0) -- (3,-1);
    \draw[red, thick ,->] (3,-1) -- (2,-2);
    \draw[blue, thick ,->] (2,0) -- (3,-1);
     \draw[blue, thick ,->]  (3,-1)--(4,-2);
         \draw[blue, thick ,->] (3,1) -- (4,0);
          \draw[blue, thick ,->]  (1,-1)--(2,-2);
           \draw[red, thick ,->] (-4,2) -- (-3,1);
    \draw[red, thick ,->] (-3,1) -- (-2,0);
    \draw[red, thick ,->] (-2,0) -- (-1,-1);
    \draw[red, thick ,->] (-1,-1) -- (-0,-2);
     \draw[red, thick ,->] (-4,0) -- (-3,-1);
    \draw[red, thick ,->] (-3,-1) -- (-2,-2);
    \draw[blue, thick ,->] (-2,0) -- (-3,-1);
     \draw[blue, thick ,->]  (-3,-1)--(-4,-2);
         \draw[blue, thick ,->] (-3,1) -- (-4,0);
          \draw[blue, thick ,->]  (-1,-1)--(-2,-2);
   \end{tikzpicture}
    \caption{The SUSY generators shift between radiative primaries corresponding to conformally soft theorems within the same helicity sector.  On the left is a zoomed in view of the action on the positive helicity sector.  This corresponds to the upper right hand corner in our top-down view of the celestial pyramid. The diamonds corresponding to the positive helicity subleading soft graviton, leading soft gravitino, and leading soft photon are highlighted.  }
    \label{fig:shiftingdiamonds}
\end{figure}
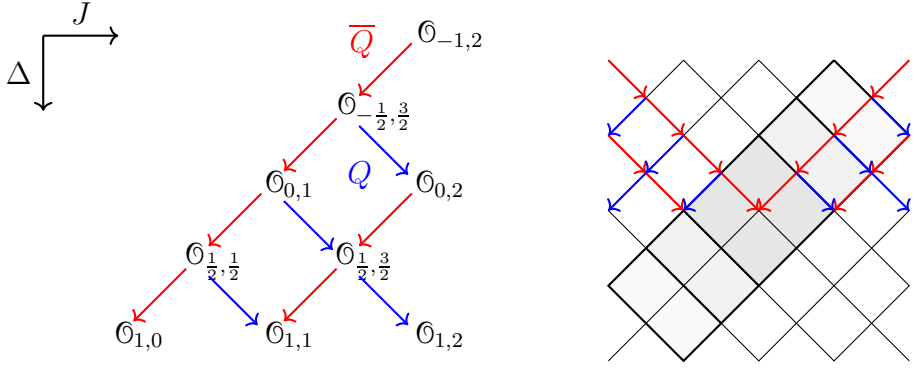

The starting point is the observation~\cite{Fotopoulos:2020bqj} that the SUSY charges map conformally soft theorems to one another. For radiative primaries (i.e. $s=|J|$) they found the following sequences of soft limits for the gauge multiplet
\be
{\cal O}_{0,1}
\xrightarrow[]{\overline{Q}} {\cal O}_{\frac{1}{2},\frac{1}{2}}\xrightarrow[] { Q}{
\cal O}_{1,1} \,,
\ee
and for the gravity multiplet
\be
{\cal O}_{-1,2}\xrightarrow[]{\overline{Q}}{\cal O}_{-\frac{1}{2},\frac{3}{2}}\xrightarrow[]{{Q}} {\cal O}_{0,2}\xrightarrow[]{\overline{Q}}{\cal O}_{\frac{1}{2},\frac{3}{2}}\xrightarrow[]{{Q}} {\cal O}_{1,2}\,.
\ee
These are summarized in a more geometric manner in figure~\ref{fig:shiftingdiamonds}, where we see that they are glued together by addition sequences starting from the fermionic primaries (which will require $\mathcal{N}>1$ SUSY) .  In the Mellin basis, the $\mathcal{N}=1$ supercharges have the following representation 
\be\label{supercharges}
Q=\p_\theta |q\rangle e^{\p_\Delta/2},~~~\overline{Q}=\theta| q] e^{\p_\Delta/2}.
\ee

We would now like to explore how these supercharges interact with our descendancy relations. From~\eqref{supercharges} we quickly see that 
\be\label{0der}
[\p_\bw,{Q}]=[\p_w,\overline{{Q}}]=0.
\ee
These observations follow from the $\mathcal{N}=1$ super-Poincar\'e algebra which adds the (anti)-commutation relations 
\be
\badat{3}
\{\overline{Q}_a,{Q}_{\dot b}\}=\sigma^\mu_{a\dot b}P_\mu,~~~[M^{\mu\nu},\overline Q_a]=\frac{1}{2}(\sigma^{\mu\nu})_a^{~b}\overline Q_b.
\eadat
\ee
In terms of the SL$(2,\mathbb{C})$ indexing, we have the global part of the $\mathfrak{sbms}_4$ algebra, namely 
\be\badat{3}
\{G_m,\bar{G}_n\}=P_{m,n},~~~[L_m,G_n]=(\frac{1}{2}m-&n)G_{m+n},~~~[\bar{L}_m,\bar{G}_n]=(\frac{1}{2}m-n)\bar{G}_{m+n}\,,
\eadat\ee
in addition to the bosonic $\mathfrak{bms}_4$ subalgebra
\be\badat{3}
  \label{BMSalg}
[L_m,L_n]=(m-n)L_{m+n},~~~&[\bar{L}_m,\bar{L}_n]=(m-n)\bar{L}_{m+n}\,,\\
[L_{n},P_{k,l}]=(\frac{1}{2}n-k)P_{k+n,l},~~~&[\bar{L}_{n},P_{k,l}]=(\frac{1}{2}n-l)P_{k,l+n}\,,\\
\eadat\ee
restricted to $\{P_{\pm\frac{1}{2},\pm\frac{1}{2}}, G_{\pm\frac{1}{2}},\bar{G}_{\pm\frac{1}{2}},L_{i},\bar{L}_i\}$.
The observation~\eqref{0der} is just the relation 
\be\label{comm} [L_{m},\bar{G}_n]=[\bar{L}_m,G_n]=0.
\ee
Our question about how the supercharges interact with the SL$(2,\mathbb{C})$ multiplets is simply a question about descendancy relations for a larger algebra.

The conditions for Poincar\'e primaries follows from a relaxation of the BMS primary construction in~\cite{Banerjee:2020kaa} for the Mellin transformed massless amplitudes.  See also~\cite{Stieberger:2018onx,Law:2019glh,Law:2020xcf}. A Poincar\'e primary is annihilated by 
\be
L_1,\bar{L}_1,P_{\frac{1}{2},\frac{1}{2}},P_{\frac{1}{2},-\frac{1}{2}},P_{-\frac{1}{2},\frac{1}{2}}\,,
\ee
and has weights $h,\bar{h}$ under $L_0,\bar{L}_0$ 
\be
L_0|h,\bar{h}\rangle=h|h,\bar{h}\rangle,~~~\bar{L}_0|h,\bar{h}\rangle=\bar{h}|h,\bar{h}\rangle\,,
\ee
while descendants are generated by
\be\label{gen1}
L_{-1},\bar{L}_{-1},P_{-\frac{1}{2},-\frac{1}{2}}\,.
\ee
Analogously, a massless $\mathcal{N}=1$ super-Poincar\'e primary is annihilated by
\be
L_1,\bar{L}_1,P_{\frac{1}{2},\frac{1}{2}},P_{\frac{1}{2},-\frac{1}{2}},P_{-\frac{1}{2},\frac{1}{2}},G_{\frac{1}{2}},\bar{G}_{\frac{1}{2}}\,,
\ee
and has weights $h,\bar{h}$ under $L_0,\bar{L}_0$, while descendants are generated by
\be\label{gen}
L_{-1},\bar{L}_{-1},G_{-\frac{1}{2}},\bar{G}_{-\frac{1}{2}}\,,
\ee
where we have used 
$\{G_{-\frac{1}{2}},\bar{G}_{-\frac{1}{2}}\}=P_{-\frac{1}{2},-\frac{1}{2}}$ to remove the translation in~\eqref{gen1} from our list. Here we will focus on the anti-chiral subalgebra spanned by $\bar{L}_i,\bar{G}_{\pm\frac{1}{2}}$, since this is enough to connect the fermionic soft charges examined in this paper to their supersymmetrically related bosonic counterparts. 
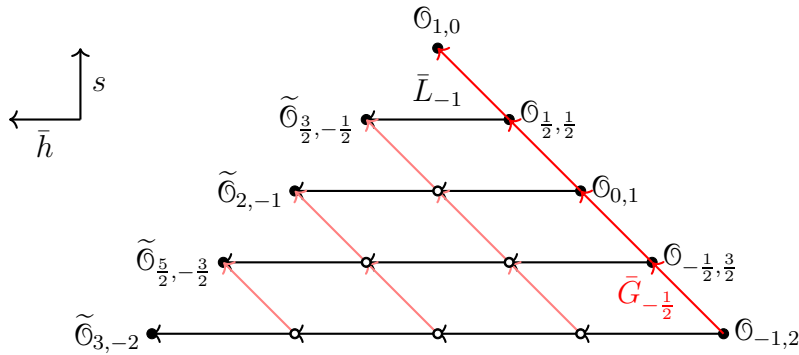
\begin{figure}[b!]
    \centering
 \begin{tikzpicture}[scale=0.95]
      \draw[thick,->] (-3,-1) -- node[right]{$s$} (-3,0) ;
     \draw[thick,->] (-3,-1) -- node[below]{$\bar{h}$} (-4,-1) ;
     \node at (4,-1) {};
\draw[thick,<-] (-2,-4) node[left]{$\widetilde{\mathcal{O}}_{3,-2}$} --(0,-4) ;
\draw[thick,<-] (0,-4)--(2,-4);
\draw[thick,<-] (2,-4)--(4,-4);
\draw[thick,<-] (4,-4)--(6,-4)node[right]{$\mathcal{O}_{-1,2}$};
\draw[thick,<-] (-1,-3) node[left]{$\widetilde{\mathcal{O}}_{\frac{5}{2},-\frac{3}{2}}$} --(1,-3);
\draw[thick,<-] (1,-3)--(3,-3);
\draw[thick,<-](3,-3)--(5,-3) node[right]{$\mathcal{O}_{-\frac{1}{2},\frac{3}{2}}$};
\draw[thick,<-] (0,-2) node[left]{$\widetilde{\mathcal{O}}_{2,-1}$} --  (2,-2) ;
\draw[thick,<-] (2,-2) --  (4,-2)node[right]{$\mathcal{O}_{0,1}$} ;
\draw[thick,<-] (1,-1) node[left] {$\widetilde{\mathcal{O}}_{\frac{3}{2},-\frac{1}{2}}$} --node[above]{$\bar{L}_{-1}$} (3,-1) node[right] {$\mathcal{O}_{\frac{1}{2},\frac{1}{2}}$} ;
\filldraw[] (2,0) circle (2pt) node[above]{$\mathcal{O}_{1,0}$};
\filldraw[] (-1,-3) circle (2pt);
\filldraw[] (0,-2) circle (2pt);
\filldraw[] (1,-1) circle (2pt);
\filldraw[] (-2,-4) circle (2pt);
\filldraw[] (6,-4) circle (2pt);
\filldraw[] (5,-3) circle (2pt);
\filldraw[] (4,-2) circle (2pt);
\filldraw[] (3,-1) circle (2pt);
\draw[thick,red,->] (6,-4) -- node[left]{$\bar{G}_{-\frac{1}{2}}$} (5,-3) ;
\draw[thick,red,->] (5,-3) -- (4,-2) ;
\draw[thick,red,->] (4,-2) -- (3,-1) ;
\draw[thick,red,->] (3,-1) -- (2,0) ;
\draw[thick,white!50!red,->] (4,-4) -- (3,-3) ;
\draw[thick,white!50!red,->] (3,-3) -- (2,-2) ;
\draw[thick,white!50!red,->] (2,-2) -- (1,-1) ;
\draw[thick,white!50!red,->] (2,-4) -- (1,-3) ;
\draw[thick,white!50!red,->] (1,-3) -- (0,-2) ;
\draw[thick,white!50!red,->] (0,-4) -- (-1,-3) ;
\filldraw[] (4,-4) circle (2pt);
\filldraw[white] (4,-4) circle (1pt);
\filldraw[] (2,-4) circle (2pt);
\filldraw[white] (2,-4) circle (1pt);
\filldraw[] (0,-4) circle (2pt);
\filldraw[white] (0,-4) circle (1pt);
\filldraw[] (2,-2) circle (2pt);
\filldraw[white] (2,-2) circle (1pt);
\filldraw[] (3,-3) circle (2pt);
\filldraw[white] (3,-3) circle (1pt);
\filldraw[] (1,-3) circle (2pt);
\filldraw[white] (1,-3) circle (1pt);
 \end{tikzpicture}
    \caption{Vertical section of the celestial pyramid corresponding to the $\Delta=1-J$ diagonal of figure~\ref{fig:shiftingdiamonds}.  The commuting generators $\bar{G}_{-\frac{1}{2}}$ and $\bar{L}_{-1}$ map between states in the positive helicity degenerate celestial diamonds corresponding to the most subleading soft theorems.   More generally, the $\Delta=k+1-J$ sections connect radiative $J=+s$ primaries of different spins and their descendant $\mathcal{O}_{soft}$. }
    \label{fig:shiftingdiamonds2}
\end{figure}

Starting from a conformal primary $\mathcal{O}_{\Delta,J}$ annihilated by $\bar{L}_{+1}$ and $\bar{G}_{+\frac{1}{2}}$
\be
\bar{L}_{1}\mathcal{O}_{\Delta,J}(0,0)=0,~~~\bar{G}_{\frac{1}{2}}\mathcal{O}_{\Delta,J}(0,0)=0
\ee
we have the $\N=1$ doublet
\be\label{doublet}
\left(\mathcal{O}_{\Delta,J},\bar{G}_{-\frac{1}{2}}\mathcal{O}_{\Delta,J}\right)\,,
\ee
where we will suppress the $(w,\bw)=(0,0)$ coordinate here and in what follows.
In terms of
\be
h=\frac{1}{2}(\Delta+J),~~~\bar{h}=\frac{1}{2}(\Delta-J)\,,
\ee
we see that the second operator in the doublet has weights $(h',\bar{h}')=(h,\bar{h}+\frac{1}{2})$.  Since
\be
\bar{L}_{1}\bar{G}_{-\frac{1}{2}}\mathcal{O}_{\Delta,J}=[\bar{L}_{1},\bar{G}_{-\frac{1}{2}}]\mathcal{O}_{\Delta,J}=\bar{G}_{\frac{1}{2}}\mathcal{O}_{\Delta,J}=0\,,
\ee
this is also an ${\mathrm{SL}}(2,\mathbb{C})$ primary.
The operators corresponding to positive helicity soft theorems have $J=+s$ and $\Delta\in\{1-s,...,s+1\}\cap(-\infty,1]$, which we can label by the value $\bar{k}\in\mathbb{Z}$
\be
\bar{k}=1+J-\Delta\,,
\ee
the level of their $\bar{L}_{-1}$ primary descendant.  Meanwhile, the superpartner $\bar{G}_{-\frac{1}{2}}\mathcal{O}_{\Delta,J}$ will have a primary descendant at level $\bar{k}'=\bar{k}-1$.  

This is consistent with the the structure of spin-$s$ celestial diamonds. The observation that~\cite{Fotopoulos:2020bqj} 
\be\label{susymult}
\bar{G}_{-\frac{1}{2}}\mathcal{O}^s_{\Delta,s}=\mathcal{O}^{s-\frac{1}{2}}_{\Delta+\frac{1}{2},s-\frac{1}{2}},
\ee
for an appropriate normalization of the operators, can be phrased in terms of the soft operators $\O^s_{soft}$ residing at the bottom corner of the spin-$s$ diamonds. From section~3 of~\cite{Pasterski:2021fjn} we have the following descendancy relations
\begin{equation}
    \O^s_{soft}=\frac{1}{\bar k!} \partial_\bw^{\bar k} \O^s_{\Delta,s}\,,\quad \mathcal{O}^{s-\frac{1}{2}}_{soft} =\frac{1}{(\bar k-1)!} \partial_\bw^{(\bar k-1)} \mathcal{O}^{s-\frac{1}{2}}_{\Delta+\frac{1}{2},s-\frac{1}{2}}\,.
\end{equation}
Using~\eqref{BMSalg}, this implies that the soft charges are related via
\be\label{softop}
\bar k \,\bar{G}_{-\frac{1}{2}}\mathcal{O}^s_{soft}=\bar{L}_{-1}\mathcal{O}^{s-\frac{1}{2}}_{soft}.
\ee
Moreover, from the celestial diamond perspective, it is clear that the observation~\eqref{symcurr} about the relation between the symmetry parameters of different spin-$s$ asymptotic symmetry currents extends to general $\bar k$. This is because the shadows of both of the operators $\left(\mathcal{O}^s_{\Delta,s},\mathcal{O}^{s-\frac{1}{2}}_{\Delta+\frac{1}{2},s-\frac{1}{2}}\right)$ appearing in the doublet~\eqref{doublet}, namely
\be
\left(\widetilde{\mathcal{O}}^s_{2-\Delta,-s},\widetilde{\mathcal{O}}^{s-\frac{1}{2}}_{\frac{3}{2}-\Delta,\frac{1}{2}-s}\right)\,,
\ee
are both level $k=k'=\Delta+s-1$ ascendants of the corresponding soft operators.  We can further see that~\eqref{softop} together with~\eqref{comm} and the shadow/descendancy relations of~\cite{Pasterski:2021fjn} imply
\be\label{shadowmult}
\bar k \,\bar{G}_{-\frac{1}{2}}\widetilde{\mathcal{O}}^s_{2-\Delta,-s}
=-\bar{L}_{-1}\widetilde{\mathcal{O}}^{s-\frac{1}{2}}_{\frac{3}{2}-\Delta,\frac{1}{2}-s}.
\ee
This analog of~\eqref{susymult} for the shadow modes plays an important role in celestial CFT since both the stress tensor and the super-current are constructed via shadow transforms. For the particular case of $\Delta=0,s=2$, the relation~\eqref{shadowmult} which involves these two symmetry currents is a statement about how the global symmetry generators act on the representation of $\mathfrak{sbms}_4$ arising from the soft theorem, consistent with
 \be
 [\bar{G}_{-\frac{1}{2}},\bar{L}_{-2}]-\frac{1}{2} [\bar{L}_{-1} ,\bar{G}_{-\frac{3}{2}}]=0.
  \ee
  
  We thus see how the spin shifting relations of~\cite{Pasterski:2021dqe} and the celestial diamond story of~\cite{Pasterski:2021dqe,Pasterski:2021fjn} fit together.  The global SL$(2,\mathbb{C})$ descendants shift within the $(\Delta,J)$ plane while the SUSY charges translate in the transverse $s$-direction.  This structure explains the connection between the gauge parameters used to demonstrate soft theorem/Ward identity equivalences for different spins.  While we have focused on the $\mathcal{N}=1$ case connecting pairs of soft theorems, this structure can straightforwardly be extended to the $\mathcal{N}>1$ case.

\section*{Acknowledgements}

The work of Y.P. is supported by the PhD track fellowship of Ecole Polytechnique. 
The work of S.P. is supported by the Sam B. Treiman Fellowship at the Princeton Center for Theoretical Science. 
The work of A.P. is supported by the European Research Council (ERC) under the European Union’s Horizon 2020 research and innovation programme (grant agreement No 852386).

\appendix{
\section{Conventions}\label{app:Conventions}

We use the conventions of~\cite{Elvang:2013cua}. 
In the Weyl representation and using signature $-+++$ we have
\be\label{eq:gamma}
\gamma^\mu=\left(\begin{array}{cc}0&(\sigma^\mu)_{a\dot b} \\ (\bar{\sigma}^\mu)^{\dot a b} &0\end{array}\right)\,, \quad \quad \{\gamma^\mu,\gamma^\nu\}=-2\eta^{\mu\nu}\,,
\ee
where we define
\begin{equation}
    (\sigma^\mu)_{a\dot{b}}=(\mathbb{1},\sigma^i)_{a\dot{b}}\,, \quad   (\bar\sigma^\mu)^{\dot{a}{b}}=(\mathbb{1},-\sigma^i)^{\dot{a}{b}}\,,
\end{equation}
in terms of the Pauli matrices $\sigma^i$.
Undotted (dotted) indices refer to left (right) handed SL(2,$\mathbb{C}$) spinors and are raised and lowered with 
\be\label{lc2}
\varepsilon^{ab}=\varepsilon^{\dot{a}\dot{b}}=-\varepsilon_{ab}=-\varepsilon_{\dot{a}\dot{b}}=\left(\begin{array}{cc}
0&1\\-1&0\\
\end{array}\right)\,,
\ee
which obey $\varepsilon_{ab}\varepsilon^{bc}=\delta_a^{~c}$. For example, we have the spinor helicity identities
\begin{equation}
    [k|^a=\varepsilon^{ab}|k]_b\,, \quad |k\rangle^{\dot{a}}=\varepsilon^{\dot{a}\dot{b}} \langle k|_{\dot{b}}\,.
\end{equation}

\paragraph{From Weyl to Dirac and Majorana}
The discussion in the main body considers Weyl photinos and gravitinos. We have the following embedding into four-component Dirac and Rarita-Schwinger fields
\begin{equation}\label{Dirac}
    \Psi=\left(\begin{array}{c} \psi_{a} \\ \bar{\psi}^{\dot a} \end{array}\right)\,, \quad\quad \Psi_\mu=\left(\begin{array}{c} \chi_{\mu a} \\ \bar{\chi}^{\dot a}_\mu \end{array}\right)\,,
\end{equation}
of the left-handed and right-handed Weyl photinos, $\psi$ and $\bar \psi$, and the left-handed and right-handed Weyl gravitinos, $\chi_\mu$ and $\bar{\chi}_\mu$. 
The Hermitian conjugate of $\Psi$ is
\be\label{Psidagger}
\Psi^\dagger=\left( (\psi^\dagger)_{\dot a},(\bar{\psi}^\dagger)^a\right)\,.
\ee
The Dirac conjugate is 
\be\label{Psibar}
\overline{\Psi}\equiv\Psi^\dagger\beta=\left((\bar{\psi}^\dagger)^a,(\psi^\dagger)_{\dot a}\right)\,,
\ee
where 
\begin{equation}
    \beta\equiv \left(\begin{array}{cc}
    0 & \delta^{\dot a}_{~\dot b} \\
    \delta_{a}^{~b}  & 0
\end{array}\right)\,,
\end{equation}
is numerically equal to $\gamma^0$ but has different index structure.
Introducing the charge conjugation matrix
\begin{equation}\label{Cmatrix}
    \C\equiv \left(\begin{array}{cc}
    \varepsilon_{ab} & 0 \\
   0 & \varepsilon^{\dot{a}\dot{b}}
\end{array}\right)\,,
\end{equation}
the charge conjugate of the Dirac field $\Psi$ is given by
\begin{equation}\label{chargeconjugate}
    \Psi^{\rm C}\equiv \C \overline{\Psi}^{\rm T} = \left(\begin{array}{c} \varepsilon_{ab}(\bar{\psi}^\dagger)^{b} \\ \varepsilon^{\dot a \dot b}(\psi^\dagger)_{\dot b} \end{array}\right)=\left(\begin{array}{c} (\bar{\psi}^\dagger)_{a} \\ (\psi^\dagger)^{\dot a} \end{array}\right)\,.
\end{equation}
We will be interested in Majorana fields for which $\Psi^{\rm C}_{\rm M}=\Psi_{\rm M}$. The Majorana condition thus amounts to
\begin{equation}\label{MajoranaCondition}
\Psi_{\rm M}=\C \overline \Psi^{\rm T}_{\rm M}\,.
 \end{equation}
These steps can be repeated for the Rarita-Schwinger field.

\paragraph{Fermion Operator Mode Expansion}
In the momentum basis, we have the following bulk mode expansions for a spin-$\frac{1}{2}$ Majorana field
\begin{equation}
    \hat \Psi_{{\rm M}}(X)=\sum_{s=\pm}\int \frac{d^3k}{(2\pi)^3}\frac{1}{2k^0}\left[a_s u_s e^{ik\cdot X}+a_s^\dagger v_s e^{-ik\cdot X}\right]\,,
\end{equation}
and a spin-$\frac{3}{2}$ Majorana field
\begin{equation}
    \hat \Psi_{{\rm M}\mu}(X)=\sum_{s=\pm}\int \frac{d^3k}{(2\pi)^3}\frac{1}{2k^0}\left[\epsilon^{*s}_\mu a_s u_s e^{ik\cdot X}+\epsilon^s_\mu a_s^\dagger v_s e^{-ik\cdot X}\right]\,,
\end{equation}
with polarization vectors $\epsilon^\pm_\mu$, and with the anti-commutation relations
\begin{equation}
    \{a_s(\vec{k}),a_{s'}^\dagger(\vec{k}')\}=(2\pi)^3(2k^0) \delta^{(3)}(\vec{k}-\vec{k}')\delta_{ss'}\,.
\end{equation}
For massless fields we can express the four-component spinors $u_s$ and $v_s$ in terms of two-component commuting spinors. In spinor-helicity notation the outgoing anti-fermions $v_s$ with helicity $s=\pm \frac{1}{2}$ are given by
\begin{equation}
    v_+ =\left(\begin{array}{c} |k]_a\\0\end{array}\right)\,, \quad v_-=\left(\begin{array}{c} 0\\ |k\rangle^{\dot a}\end{array}\right)\,, 
\end{equation}
and the outgoing fermions $\bar u_s$  with helicity $s=\pm \frac{1}{2}$ are given by
\begin{equation}
    \overline u_-=(0,\langle k|_{\dot{a}})\,, \quad \overline u_+=([k|^a,0)\,.
\end{equation}  
Crossing symmetry flips the sign of the helicity and exchanges incoming $\leftrightarrow$ outgoing and fermion $\leftrightarrow$ antifermion, thus
\begin{equation}
    u_\mp=v_\pm\,, \quad \bar v_\mp = \bar u_\pm\,.
\end{equation}
The mode expansion of the left-handed massless Weyl photino is then
\begin{equation}\label{hatpsiexpansion}
   \hat{\psi}_a(X)=\int \frac{d^3k}{(2\pi)^3}\frac{|k]_a}{2k^0}\left[a_- e^{ik\cdot X}+a_+^\dagger e^{-ik\cdot X}\right]\,, 
\end{equation}
while for the right-handed massless Weyl photino it is
\begin{equation}\label{hatpsibarexpansion}
  \hat{\bar{\psi}}^{\dot a}(X)=\hat{\psi}^{\dagger \dot a}(X)=\int \frac{d^3k}{(2\pi)^3}\frac{| k\rangle^{\dot a}}{2k^0}\left[a_+ e^{ik\cdot X}+a_-^\dagger e^{-ik\cdot X}\right]\,.
\end{equation}
The mode expansion for the left-handed massless Weyl gravitino is
\begin{equation}\label{hatchiexpansion}
   \hat{\chi}_{\mu a}(X)=\int \frac{d^3k}{(2\pi)^3}\frac{|k]_a}{2k^0}\epsilon^+_\mu\left[ a_- e^{ik\cdot X}+a_+^\dagger e^{-ik\cdot X}\right]\,, 
\end{equation}
while for the right-handed massless Weyl gravitino it is 
\begin{equation}\label{hatchibarexpansion}
  \hat{\bar{\chi}}_\mu^{\dot a}(X)=\hat{\chi}^{\dagger \dot a}_\mu(X)=\int \frac{d^3k}{(2\pi)^3}\frac{|k\rangle^{\dot a}}{2k^0}\epsilon^-_\mu\left[a_+ e^{ik\cdot X}+a_-^\dagger e^{-ik\cdot X}\right]\,.
\end{equation}

\section{Radial Expansions of Radiative Solutions}
\label{app:radexpfree}
In this section we explore the radial expansions for standard radiative solutions to the chiral Dirac and Rarita-Schwinger equations.  This will let us identify the free data which get promoted to the mode operators seen in section~\ref{sec:OPexpansion}, as well as provide more context for the shadow versus non-shadow fall offs explored in section~\ref{sec:CPWexpansion} and appendix~\ref{app:radexpCPW} below. 

\subsection{Chiral Photino}\label{app:WeylSpinor}
The massless Dirac equation decouples into the following equations for left and right-handed Weyl spinors $\psi$ and $\bar{\psi}$, respectively
\be \label{12Weyl}
\bar{\sigma}^\mu\p_\mu \psi=0\,,~~~\sigma^\mu\p_\mu \bar{\psi}=0\,.
\ee
Here we will focus on the left-handed photinos for which we make the following ansatz for the radial expansion
\be\label{photinoansatz}
 \psi(u,r,z,\bz)=\sum_n r^{-n} \Big[ \psi^{(n)}(u,z,\bz)+\log r\, \hat{\psi}^{(n)}(u,z,\bz)\Big]\,.
\ee
For these towers to truncate and to have sensible boundary conditions we should impose the vanishing of modes $\psi^{(n)}$ and $\hat \psi^{(n)}$ starting at some values of $n$.  Introducing $\bar{D}=r\sigma^A\p_A$, whose $r$ dependence cancels, we have the compact relation
 \be\label{12Towers}
 \bar{\sigma}_r\p_u \psi^{(n)}+(\bar{\sigma}_r-\bar{\sigma}_u)\big((n-1)\psi^{(n-1)}-\hat{\psi}^{(n-1)}\big)-\bar{D}\psi^{(n-1)}=0\,.
 \ee
The same equation holds for the log modes after substituting $\hat{\psi}^{(n)}\mapsto0$ and $\psi^{(n)}\mapsto\hat{\psi}^{(n)}$.
 Meanwhile
\be
\{\sigma^\mu,\bar{\sigma}^\nu\}=-2\eta^{\mu\nu}\mathbb{1}~~\Rightarrow~~\sigma^\nu\p_\nu{\bar\sigma}^\mu\p_\mu=-2\Box,
\ee
so our solutions also satisfy the Klein-Gordon equation for each component. This implies that the radial order of the free data should match that of the massless scalar. We can thus impose $\psi^{({n\le }0)}(u,z,\bz)$ and $\hat{\psi}^{({n\le 1})}(u,z,\bz)$ vanishing, and will find constraints on the components of $\psi^{(1)}(u,z,\bz)$ which we now identify. Since
 \be\label{eq:xspin2}
\bar{\sigma}_r=-\frac{2}{1+z\bz}\left(\begin{array}{cc}1&\bz\\z&z\bz\end{array}\right)=-\frac{
1}{1+z\bz}| x\rangle^{\dot{a}} [x|^{{a}}\,,
\ee
we see that equation~\eqref{12Towers} imposes the constraint 
\be\label{GenphotinoFalloff}
\p_u\big(\psi^{\dot{1}(1)}+\bz\psi^{\dot{2}(1)}\big)=0\,. 
\ee
This enforces
\be
\psi^{(1)}\propto\left(\begin{array}{c}
\bz\\-1
\end{array}\right)\,,
\ee
such  that the free data is proportional to the spinor $|x]_a$.

\subsection{Chiral Gravitino}\label{app:chigrav}
The gravitino example has been worked out in~\cite{Lysov:2015jrs} and we will use this appendix to translate notation. There, the following gauge fixing conditions are imposed on the Dirac spinor
\be\label{Psimugaugecond}
\nabla^\mu\Psi_\mu=0\,, \quad \gamma^\mu\Psi_\mu=0\,.
\ee
It is straightforward to check that the conformal primary wavefunctions also satisfy these constraints.
Focusing here on the left-handed Weyl gravitinos, 
we have the radial expansion
\be
 \chi_\mu(u,r,z,\bz)=\sum_n r^{-n} \Big[ \chi_\mu^{(n)}(u,z,\bz)+\log r\, \hat{\chi}_\mu^{(n)}(u,z,\bz)\Big]\,.
\ee
As for the photino in~\eqref{photinoansatz}, we will not need the log tower for generic gravitino perturbations without sources. 
On the modes $\chi^{(n)}_\mu(u,z,\bz)$ we then impose fall-offs such that
\begin{equation}
{\chi}_{r}^{(0)}={\chi}_{u}^{(0)}={\chi}_{r}^{(1)}=0\,.
\ee
The Rarita-Schwinger equation together with the gauge conditions~\eqref{Psimugaugecond} impose further constraints.
The projection operators introduced in~\cite{Lysov:2015jrs} take the form
\be\scalemath{0.95}{\label{p0p1}
P_0=-\frac{1}{2}\gamma_{r}\gamma_u=-\frac{1}{2}
\left(\begin{array}{cc}\sigma_r &0 \\ 0&\bar{\sigma}_r\end{array}\right)
,~~~P_1=-\frac{1}{2}\gamma_{u}\gamma_r=-\frac{1}{2}\left(\begin{array}{cc}\bar{\sigma}_r &0 \\ 0&{\sigma}_r \end{array}\right)}\,,
\ee
where ${\bar\sigma}_r$ is given in~\eqref{eq:xspin2} and
\be\label{eq:xspin1}
\sigma_r=-\frac{2}{1+z\bz}\left(\begin{array}{cc}z\bz&-\bz\\-z&1\end{array}\right)=-\frac{1}{1+z\bz}|x]_{{a}}\langle x|_{\dot{a}}\,.
\ee
In our notation, the constraints $r\,\gamma^A P_0\Psi^{(0)}_A=0$ and $P_1\Psi^{(0)}_A=0$ in (3.11) of~\cite{Lysov:2015jrs} demand on the components of $\chi^{(0)}_z$ and $\chi^{(0)}_\bz$ that
\be
z\chi_{1z}^{(0)}-\chi_{2z}^{(0)}=0\,,\quad \text{and} \quad \chi_{1\bz}^{(0)}+\bz\chi_{2\bz}^{(0)}=0\,,~~\chi_{1z}^{(0)}+\bz\chi_{2z}^{(0)}=0\,.
\ee
This yields $\chi_{1z}^{(0)}=\chi_{2z}^{(0)}=0$ and
\be
\chi_{\bz}^{(0)}\propto \left(\begin{array}{c}\bz\\-1\end{array}\right)\,,
\ee
such that the free data is proportional to the spinor $|x]_a$.

\section{Radial Expansions of Conformal Primaries}\label{app:radexpCPW}
We detail here the leading large-$r$ data of our radiative conformal primary wavefunctions from section~\ref{sec:CPW} and use it as an opportunity to check the shadow relations for massless fermions. In 2D CFT the shadow transform of a primary operator $\O_{\Delta,J}$ is given by a $(\tDelta,\tJ)=(2-\Delta,-J)$ primary operator~\cite{Osborn:2012vt} 
\begin{equation}\label{2dShadowTransform}
\widetilde{\O}_{\tDelta,\tJ}(w,\bw)=\frac{k_{\Delta,J}}{2\pi} \int d^2w' \frac{\O_{\Delta,J}(w',\bw')}{(w-w')^{2-\Delta-J}(\bw-\bw')^{2-\Delta+J}}\,.
\end{equation}
Using the integral relation~\cite{Dolan:2011dv} 
\begin{equation}\label{I2}
 \badat{2}
 \int d^2 z \prod_{i=1}^2 \frac{1}{(z-z_i)^{h_i}}\frac{1}{(\bz-\bz_i)^{\bar h_i}}=\frac{\Gamma(1-h_1)\Gamma(1-h_2)}{\Gamma(\bar h_1)\Gamma(\bar h_2)}(-1)^{h_1-\bar h_1}(2\pi)^2 \delta^{(2)}(z_1-z_2)\,,
\eadat\end{equation}
where $h_1+h_2=\bar h_1+\bar h_2=2$, $h_i-\bar h_i \in \mathbb Z$, we see that for the normalization factor~\cite{Pasterski:2017kqt} 
\be \label{2DSHnorm}
k_{\Delta,J}=\Delta-1+|J|,
\ee 
the shadow transform~\eqref{2dShadowTransform} squares to $(-1)^{2J}$. Comparing to~\eqref{Ffermionic} we see that these shadow relations should hold at the level of the wavefunctions as well, namely $\mathcal{O}_{\Delta,J}\mapsto\Psi_{\Delta,J}$. 
We will now verify that~\eqref{psichi} and~\eqref{SHpsichi} when expanded near null infinity using the saddle point approximation~\eqref{saddlept} are indeed related by a 2D shadow transform on the celestial sphere.

\subsection{Chiral Photino}

Near future null infinity the left-handed $J=+\frac{1}{2}$ Weyl spinor has the expansion
\begin{equation}\label{lhmodespsi}\badat{2}
    \psi_{\Delta,J=+\frac{1}{2}}\Big|_{z=w}&=\sqrt{2}i\frac{\pi}{\Delta-\frac{1}{2}}(1+z\bz)^{\frac{3}{2}-\Delta}\left(
\begin{array}{c}
\bw \\ -1
\end{array}\right)u^{\frac{1}{2}-\Delta}r^{-1}\delta^{(2)}(z-w)+...\,,\\
    \psi_{\Delta,J=+\frac{1}{2}}\Big|_{z\neq w}&=\sqrt{2}i\left(\frac{1+z\bz}{2(z-w)(\bz-\bw)}\right)^{\Delta+\frac{1}{2}}\left(
\begin{array}{c}
\bw \\ -1
\end{array}\right)r^{-\Delta-\frac{1}{2}}+...\,,
\eadat
\end{equation}
while its $J=-\frac{1}{2}$ shadow transformed spinor falls off as 
\begin{equation}  \scalemath{0.95}{  \badat{2}
 \tpsi_{\Delta,J=-\frac{1}{2}}\Big|_{z=w}&=i2^{\Delta-1}\frac{\pi}{\frac{1}{2}-\Delta}(1+z\zb)^{\frac{5}{2}-\Delta}\left[\begin{pmatrix}
1\\0
\end{pmatrix}
+
\frac{1}{\frac{3}{2}-\Delta}
\begin{pmatrix}
\zb\\-1
\end{pmatrix}
\partial_{\zb}\right]r^{\Delta-\frac{5}{2}}\delta^{(2)}(z-w)+... \,,\\
    \tpsi_{\Delta,J=-\frac{1}{2}}\Big|_{z\neq w}&= 
-\frac{i}{\sqrt{2}(\bz-\bw)}\left(\frac{1+z\bz}{(z-w)(\bz-\bw)}\right)^{\Delta-\frac{1}{2}}\left(
\begin{array}{c}
\bz \\ -1
\end{array}\right)
u^{\Delta-\frac{3}{2}}r^{-1}+...\,.
\eadat}\end{equation}
Using~\eqref{2dShadowTransform} we see the shadow transform exchanges the leading contact and non-contact terms.

\subsection{Chiral Gravitino}
 For spin-$\frac{3}{2}$ we have more components to keep track of. For the purposes of verifying the shadow relations for radiative modes, it is sufficient to write down what would be the leading terms in the expansion when $-\frac{1}{2}<\mathrm{Re}(\Delta)$.  
Near future null infinity the left-handed $J=+\frac{3}{2}$ wavefunction has the expansion
\begin{equation}\label{lhmodeschi}
    \chi_{\Delta,J=+\frac{3}{2};z}=-2^{-\Delta+\frac{1}{2}}\frac{i}{(z-w)^2}\left(\frac{1+z\bz}{(z-w)(\bz-\bw)}\right)^{\Delta-\frac{1}{2}}\left(
\begin{array}{c}
\bw \\ -1
\end{array}\right)r^{-\Delta+\frac{1}{2}}+...\,,\\
\end{equation}
and
\begin{equation}\badat{3}
    \chi_{\Delta,J=+\frac{3}{2};\bz}&=2i\frac{\pi}{\Delta+\frac{1}{2}}(1+z\bz)^{\frac{1}{2}-\Delta}\left(
\begin{array}{c}
\bw \\ -1
\end{array}\right)u^{\frac{1}{2}-\Delta}\delta^{(2)}(z-w)+...\,,\\
\eadat\end{equation}
for the angular components while the temporal and radial components are subleading. Similarly, the angular components of the shadow transformed $J=-\frac{3}{2}$ wavefunctions have the expansion  
\begin{equation}
    \tchi_{\Delta,J=-\frac{3}{2};z}=  2^{\Delta-\frac{1}{2}}i\frac{\pi}{\Delta+\frac{1}{2}}(1+z\bz)^{\frac{3}{2}-\Delta}\Big[\left(
\begin{array}{c}
1 \\ 0
\end{array}\right)+\frac{1}{\frac{1}{2}-\Delta}\left(
\begin{array}{c}
\bz \\ -1
\end{array}\right)\p_\bz\Big]r^{\Delta-\frac{3}{2}}
 \delta^{(2)}(z-w)+...\,,
\end{equation}
and
\begin{equation}
    \tchi_{\Delta,J=-\frac{3}{2};\bz}=
\frac{-i}{(\bz-\bw)^3}\left(\frac{1+z\bz}{(z-w)(\bz-\bw)}\right)^{\Delta-\frac{3}{2}}\left(
\begin{array}{c}
\bz \\ -1
\end{array}\right)
u^{\Delta-\frac{3}{2}}+...\,.
\end{equation}
Using~\eqref{2dShadowTransform} we can again verify the shadow relations for all radiative gravitinos, as promised in~\cite{Pasterski:2020pdk}.  The expansions above also suffice for computing the charge for the leading conformally soft gravitino at $\Delta=\frac{1}{2}$, or equivalently its shadow at $\Delta=\frac{3}{2}$.  For the subleading soft gravitino at $\Delta=-\frac{1}{2}$, or its shadow at $\Delta=3$, some of the suppressed terms in the above expansions become the same order as the ones shown. We will revisit this in section~\ref{sec:subgravitino} where we compute analogous soft charges. 

\section{Inner Products \texorpdfstring{($\Psi_{\Delta,J},\Psi'_{\Delta',J'}$)}{}}\label{app:IP}

As an alternative to the momentum basis mode expansions used in the main text, we can follow~\cite{Donnay:2020guq} and expand the free spin-$s$ Heisenberg picture operators in a conformal primary basis.  In terms of the chiral Dirac spinors~\eqref{chiralspinor}, we have
\be\label{BulkOexpansion}
\begin{array}{rl}
O^{s}(X^\mu)=\sum\limits_{J=\pm s}\int d^2 w \int_{1-i\infty}^{1+i\infty}(-id\Delta)& \Big[{\cal N}^+_{2-\Delta,s}{\Psi}^{s}_{2-\Delta,-J}(X_+^\mu;w,\bw)b_{\Delta,J}(w,\bw)\\
&~+{\cal N}^-_{\Delta,s}\Psi^s_{\Delta,J}(X_-^\mu;w,\bw)b_{\Delta,J}(w,\bw)^\dagger\Big]\,.
\end{array}
\ee
In this appendix we  will evaluate the inner products for spins $\frac{1}{2}$ and $\frac{3}{2}$ on a spacelike Cauchy slice to find
\be
(\Psi^s_{\Delta,J}(X_\pm),\Psi'^s_{\Delta',J'}(X_\pm))=\pm \left(\N^\pm_{\Delta,s}\right)^{-2}\delta_{JJ'}\delta^{(2)}(w-w')\boldsymbol{\delta}(i(\Delta+\Delta'-2))\,,
\ee
where the coefficients
\begin{equation}\label{N012}
{\textstyle
\mathcal{N}^\pm_{\Delta,\frac{1}{2}}=\left[
\frac{(2\pi)^4e^{\pm i \pi \Delta}\cos(\pi\Delta)}{\pi(\Delta-\frac{1}{2}){(\frac{3}{2}-\Delta)}}\right]^{-\frac{1}{2}}, \; \mathcal{N}^\pm_{\Delta,\frac{3}{2}}=\left[
\frac{(2\pi)^4e^{\pm i \pi \Delta}\cos(\pi\Delta)}{\pi(\Delta-\frac{1}{2}){(\frac{3}{2}-\Delta)}}\right]^{-\frac{1}{2}}}\,,
\end{equation}
are designed to give the canonical commutation relations 
\be
\label{commutation_modes}
\{b_{\Delta,J}(w,\bw),b_{\Delta',J'}(w',\bw')^\dagger\}=\delta_{JJ'}\delta^{(2)}(w-w')\boldsymbol{\delta}(i(\Delta+\Delta'^*-2))\,.
\ee
The generalized distribution $\boldsymbol{\delta}$ which was defined in~\cite{Donnay:2020guq} reduces to the ordinary Dirac delta function for $\Delta,\Delta'\in 1+i\mathbb{R}$ on the principal series.

\subsection*{Spin-\texorpdfstring{$\frac{1}{2}$}{}}

Starting from the spin-$\frac{1}{2}$ inner product~\cite{Muck:2020wtx}\footnote{
 Because we want an inner product on positive frequency solutions, this is essentially a complexification of $-i\Omega(\Psi,\Psi')$, where $\Omega$ is the symplectic product, and the $\Psi$ are Majorana. The symplectic product will be the focus of the next section. 
}
\be\label{IPphotino}
(\Psi,\Psi')=\int d\Sigma^\mu \bar{\Psi}'\gamma_\mu \Psi\,,
\ee
and using a constant time Cauchy slice,
we see that for $J=+\frac{1}{2}$ we have
\be\badat{3}
(\Psi^{\pm}_{\Delta,+\frac{1}{2}},\Psi^{\pm}_{\Delta',+\frac{1}{2}})&=\langle \pm q'|_{{\dot{a}}}\bar{\sigma}^{0\dot{a}a}|\pm q]_a\int d^3 X \frac{1}{(-q\cdot X_\pm )^{\Delta+\frac{1}{2}}(-q'\cdot X_\mp )^{\Delta'^*+\frac{1}{2}}}\,,\\
\eadat\ee
while for $J=-\frac{1}{2}$
\be\badat{3}
(\Psi^{\pm}_{\Delta,-\frac{1}{2}},\Psi^{\pm}_{\Delta',-\frac{1}{2}})&=[ \pm q'|^a{\sigma}^{0}_{a\dot{a}}|\pm q\rangle^{\dot a}\int d^3 X \frac{1}{(-q\cdot X_\pm )^{\Delta+\frac{1}{2}}(-q'\cdot X_\mp )^{\Delta'^*+\frac{1}{2}}}\,,\\
\eadat\ee
giving
\be
(\Psi_{\Delta,J}(X_\pm),\Psi'_{\Delta',J'}(X_\pm))=\pm 
\frac{(2\pi)^4e^{\pm i \pi \Delta}\cos(\pi\Delta)}{\pi(\Delta-\frac{1}{2})(\frac{3}{2}-\Delta)} \delta_{JJ'}\delta^{(2)}(w-w')\boldsymbol{\delta}(i(\Delta+\Delta'^*-2))\,,
\ee
so that
\be
{
\mathcal{N}^\pm_{\Delta,\frac{1}{2}}=\left[
\frac{(2\pi)^4e^{\pm i \pi \Delta}\cos(\pi\Delta)}{\pi(\Delta-\frac{1}{2})(\frac{3}{2}-\Delta)}\right]^{-\frac{1}{2}}}.
\ee

\subsection*{Spin-\texorpdfstring{$\frac{3}{2}$}{}}
For spin-$\frac{3}{2}$ we start from 
\be\label{IPgravitino}
(\Psi,\Psi')=\int d\Sigma^\mu \bar{\Psi}'^\nu \gamma_\mu \Psi_\nu\,,
\ee
and again use a constant time Cauchy slice. We see that for $J=+\frac{3}{2}$ we have
\be\badat{3}
(\Psi^{\pm}_{\Delta,+\frac{3}{2}},\Psi^{\pm}_{\Delta',+\frac{3}{2}})&=\langle \pm q'|_{{\dot a}}\bar{\sigma}^{0\dot{a}a}|\pm q]_a\int d^3 X \frac{m\cdot {\bar{m}}'}{(-q\cdot X_\pm )^{\Delta+\frac{1}{2}}(-q'\cdot X_\mp )^{\Delta'^*+\frac{1}{2}}}\,,\\
\eadat\ee
while for $J=-\frac{3}{2}$
\be\badat{3}
(\Psi^{\pm}_{\Delta,-\frac{3}{2}},\Psi^{\pm}_{\Delta',-\frac{3}{2}})&=[ \pm q'|^a{\sigma}^{0}_{a\dot{a}}|\pm q\rangle^{\dot a}\int d^3 X \frac{{\bar{m}}\cdot m'}{(-q\cdot X_\pm )^{\Delta+\frac{1}{2}}(-q'\cdot X_\mp )^{\Delta'^*+\frac{1}{2}}}.\\
\eadat\ee
Then using the fact
\be
m_\mu\frac{1}{(-q\cdot X)^\Delta}=\left[\epsilon_{+\mu}+{\frac{1}{\sqrt{2}\Delta}q_\mu\p_w}\right]\frac{1}{(-q\cdot X)^\Delta}\,,
\ee
and following the same manipulations as in appendix A of \cite{Donnay:2020guq}, which in this case allows us to drop the terms other that $\epsilon_+\cdot\epsilon_-'$, we find
\be
(\Psi_{\Delta,J}(X_\pm),\Psi'_{\Delta',J'}(X_\pm))=\pm 
\frac{(2\pi)^4e^{\pm i \pi \Delta}\cos(\pi\Delta)}{\pi(\Delta-\frac{1}{2})(\frac{3}{2}-\Delta)} \delta_{JJ'}\delta^{(2)}(w-w')\boldsymbol{\delta}(i(\Delta+\Delta'^*-2))\,,
\ee
so that
\be
\mathcal{N}^\pm_{\Delta,\frac{3}{2}}=\left[
\frac{(2\pi)^4e^{\pm i \pi \Delta}\cos(\pi\Delta)}{\pi(\Delta-\frac{1}{2})(\frac{3}{2}-\Delta)}
\right]^{-\frac{1}{2}}.
\ee

\section{Symplectic Structure \texorpdfstring{$\Omega(\delta \Psi,\delta'\Psi)$}{}}\label{app:Omega}

Starting from the action for Grassmann-valued spin $s=\frac{1}{2}$ and $\frac{3}{2}$ Dirac fields\footnote{Note that our gauge fixing brings the Rarita-Schwinger action to the form~\eqref{action}, and we omit there the spacetime indices in the contraction between the gravitinos.}
\begin{equation}\label{action}
S=i\int d^4X\, \bar{\Psi}{^s}\slashed{\partial}\Psi{^s}\,,
\end{equation}
where $\slashed{\p}\equiv \gamma^\mu \partial_\mu$, we compute the symplectic structure 
\begin{equation}\label{Omega}
 \Omega(\delta \Psi^s,\delta'\Psi^s)=\int_{\I^+} du d^2z  \,\omega(\delta \Psi^s,\delta'\Psi^s)\,,
\end{equation}
for the Grassmann-valued perturbations $\delta \Psi{^s}$ and $\delta'\Psi{^s}$. In \eqref{Omega}, the presymplectic form $\omega(\delta \Psi^s,\delta'\Psi^s)$ is a codimension-1 form which we integrate on a spatial slice pushed to future null infinity $\scri^+$. 
The null normal to $\scri^+$ is $n^\mu\p_\mu=\p_u-\frac{1}{2}\p_r$ and so the integrand of~\eqref{Omega} becomes $-(\frac{1}{2}\omega^u+\omega^r)$.
In this appendix, we evaluate $\Omega(\delta \Psi^s,\delta'\Psi^s)$ for $s=\frac{1}{2},\frac{3}{2}$ to obtain the soft charges related to the leading soft photino and gravitino theorems and to the subleading soft gravitino theorem. 
For the Majorana spinors we consider here we have $\overline{\Psi}_{\rm M}=\Psi^{\rm T}_{\rm M}\C$ and the Majorana flip relation 
\begin{equation}
   \delta' \overline{\Psi}_{\rm M}\gamma_\mu \delta\Psi_{\rm M}=-\delta\overline{\Psi}_{\rm M} \gamma_\mu \delta'\Psi_{\rm M}\,.
\end{equation}
The appropriate normalization of the kinetic term thus differs by a factor of $\frac{1}{2}$ from the Dirac case.

\subsection*{Spin-\texorpdfstring{$\frac{1}{2}$}{}}
Using the conventions in appendix \ref{app:Conventions}, the presymplectic structure can be written as
\be\label{omegagauino}
\omega^\nu(\delta \psi,\delta'\psi)={-\frac{i}{2}}r^2\sqrt{\gamma}\left(\delta'\psi^{\dagger}_{\dot{a}}\sb^{\nu\, \dot{a}b}\delta\psi_{ b}+\delta'\bar\psi^{\dagger\, a}\sigma^\nu_{a\dot{b}}\delta\bar\psi^{\dot{b}}-(\delta'\leftrightarrow \delta)\right)\, .
\ee
Using the Majorana condition \eqref{MajoranaCondition}, and the Majorana flip relation we can write~\eqref{omegagauino} as $\omega=\f+\f^\dagger$, we can focus on calculating
\be
\f^\nu(\delta\psi,\delta'\psi)={i}r^2\sqrt{\gamma}\delta\psi^{\dagger}_{\dot{a}}\sb^{\nu\, \dot{a}b}\delta'\psi_{ b}\,.
\ee
We consider the following fall-offs for the field variations
\begin{equation}
  \delta \psi=\frac{1}{r}\delta \psi^{(1)}(u,z,\bz)  + \O(1/r^2)\,,
\end{equation}
and
\begin{equation}
 \delta'\psi=\frac{1}{r} \delta'\psi^{(1)}(u,z,\bz) + \O(1/r^2)\,.
\end{equation}
Therefore, we find 
\begin{equation}
\badat{2}
 \f^u&={-i}\sqrt{\gamma}\left[\delta{\psi}^{\dagger(1)}\sb_r\delta'\psi^{(1)}\right]+\mathcal{O}(1/r)\,,\\
 \f^r&={i}\sqrt{\gamma}\left[\delta{\psi}^{\dagger(1)}(\sb_r-\sb_u)\delta'\psi^{(1)}\right]+\mathcal{O}(1/r)\,.
\eadat
\end{equation}
Using \eqref{GenphotinoFalloff} and noticing that 
\be\label{Simp}
\begin{pmatrix}
z & -1
\end{pmatrix}
\sb_r=\sb_r\begin{pmatrix}
\bz \\ -1
\end{pmatrix}=0\,,
\ee
we can see that $\f^u$ becomes subleading and the expression of $\omega^r$ is reduced.
This yields the symplectic structure
\begin{equation}\label{Omegapsi}
  \Omega(\delta \psi,\delta'\psi)={i}\int_{\I^+} du d^2z \sqrt{\gamma}
 \left[\delta{\psi}^{\dagger(1)} \sb_u\delta'\psi^{(1)}\right] +h.c. \,.
\end{equation}
In section~\ref{sec:photino} we are interested in the case where $\delta'\psi$ is replaced by the conformally soft photino primary.

\subsection*{Spin-\texorpdfstring{$\frac{3}{2}$}{}}
The presymplectic form for the gravitino field can be written as
\be
\omega^\nu(\delta \chi,\delta'\chi)=-{\frac{i}{2}}r^2\sqrt{\gamma}\left(\delta'\chi^{\dagger\, \mu}_{\dot{a}}\sb^{\nu\, \dot{a}b}\delta\chi_{ b\, \mu}+\delta'\bar\chi^{\dagger\, a\, \mu}\sigma^\nu_{a\dot{b}}\delta\bar\chi^{\dot{b}}_\mu-(\delta'\leftrightarrow \delta)\right)\, .
\ee
Using similar arguments as for the photino presymplectic form, we can focus on 
\be
\f^\nu(\delta\chi,\delta'\chi)={i}r^2\sqrt{\gamma}\delta\chi^{\dagger\, \mu}_{\dot{a}}\sb^{\nu\, \dot{a}b}\delta'\chi_{ b\, \mu}\,.
\ee
We consider the following fall-offs for the generic gravitino field\footnote{Notice that $\delta\chi^{(0)}_\bz\neq 0$ while $\delta\chi^{\dagger(0)}_z\neq0$ since $\epsilon^+_\bz=\epsilon^-_z\neq0$ at first order in the saddle point approximation.}
\begin{equation}\label{GenGravFalloffs}
 \begin{aligned}
  \delta \chi_{\zb}&=\delta \chi_{\zb}^{(0)}(u,z,\bz)  + \frac{1}{r} \delta \chi_{\zb}^{(1)}(u,z,\bz) +\O(1/r^2)\,,\\
  \delta \chi_z&=\frac{1}{r}\delta\chi_z^{(1)}(u,z,\bz)+\mathcal{O}(1/r^2)\,,\\
  \delta \chi_u&=\frac{1}{r}\delta\chi_u^{(1)}(u,z,\bz)+\mathcal{O}(1/r^2)\,,\\
  \delta \chi_r&=\frac{1}{r^2}\delta\chi_r^{(2)}(u,z,\bz)+\mathcal{O}(1/r^3)\, .
\end{aligned}
\end{equation}
For the fall-offs of $\delta'\chi$, we are interested in two cases: one that is compatible with the fall-offs of the leading soft gravitino and another with that of the subleading soft gravitino.

\paragraph{Leading} For the leading case, we consider 
\begin{equation}\label{chioverleading1}
\badat{2}
 \delta'\chi_{A}&=\delta'\chi_{A}^{(0)}(z,\bz) + \O(1/r)\,.
 \eadat
\end{equation}
The radial and temporal components are subleading and do not contribute to the presymplectic form. We get
\begin{equation}
\scalemath{1}{
\label{presympGravitino}
\badat{2}
 \f^u&= {-i}\left[\delta{\chi}^{\dagger(0)}_{z}\, \sb_r\, \delta'\chi^{(0)}_{\bz}\right]+\mathcal{O}(1/r)\,,\\
 \f^r&={i}\left[\delta{\chi}^{\dagger(0)}_{z}\, (\sb_r-\sb_u)\, \delta'\chi^{(0)}_{\bz}\right]+\mathcal{O}(1/r)\,.
\eadat
}
\end{equation}
Using \eqref{Simp}, this simplifies to
\be
\begin{aligned}
\f^u&=\O(1/r)\,,\\
\f^r&= {-i}\delta{\chi}^{\dagger(0)}_{z}\, \sb_u\, \delta'\chi^{(0)}_{\bz}+\mathcal{O}(1/r)\,.
\end{aligned}
\ee
This yields the symplectic structure for the leading soft gravitino 
\begin{equation}\label{Omegaleadingchi}
    \Omega(\delta \chi,\delta'\chi)=
   {i}\int_{\I^+} du d^2z \left[\delta \chi^{\dagger(0)}_z {\textstyle\sb_u} \delta'\chi^{(0)}_\bz\right]+h.c.\,.
\end{equation}
In section~\ref{sec:gravitinocharge} we replace $\delta'\chi$ by the conformally soft gravitino primary to obtain the leading soft charge for spontaneously broken large supersymmetry.

\paragraph{Subleading} For the subleading soft gravitino, we consider the following fall-offs
\begin{equation}\label{chioverleading}
\badat{2}
 \delta'\chi_{\zb}&=\delta'\chi_{\zb}^{(0)}(u,z,\bz) + \O(1/r)\,,\\
 \delta'\chi_{z}&=r\,\delta'\chi_z^{(-1)}(z,\bz)+\delta'\chi_z^{(0)}(u,z,\zb)+\O(1/r)\,,\\
 \delta'\chi_{u}&=\delta'\chi_u^{(0)}(u,z,\bz)+\mathcal{O}(1/r)\,,\\
  \delta'\chi_{r}&=\frac{1}{r}\delta'\chi_r^{(1)}(u,z,\bz)+\mathcal{O}(1/r^2)\,.
 \eadat
\end{equation}
Notice that there will be contributions from the $u$ and $r$ components of the gravitino field. Using \eqref{Simp}, the presymplectic form can be written as
\be
\begin{aligned}
\f^u&={i}\sqrt{\gamma}\left[\delta\chi^{\dagger\, (2)}_r\sb_r\delta'\chi^{(0)}_u+\delta\chi^{\dagger\, (1)}_u\sb_r\delta'\chi^{(1)}_r\right]\\
&\qquad {-i}\delta\chi^{\dagger\, (1)}_\bz \sb_r\delta'\chi^{(-1)}_z+\O(1/r)\,,\\[10pt]
\f^r&= {-i}\sqrt{\gamma}\left[\delta\chi^{\dagger\, (2)}_r(\sb_r-\sb_u)\delta'\chi^{(0)}_u+\delta\chi^{\dagger\, (1)}_u(\sb_r-\sb_u)\delta'\chi^{(1)}_r\right]\\
&\qquad +{i}\delta\chi^{\dagger\, (1)}_\bz (\sb_r-\sb_u)\delta'\chi^{(-1)}_z-i\delta\chi_z^{\dagger\, (0)}\sb_u\delta\chi^{(0)}_\bz+\O(1/r)\,.
\end{aligned}
\ee
 Looking at \eqref{chioverleading}, we notice an overleading behavior in $\delta'\chi_z$, however the overall contribution to the presymplectic form is finite because we can see from \eqref{GenGravFalloffs} that $\delta\chi^\dagger_\bz$ is subleading. The fact that $\epsilon^+_\bz=\epsilon^-_z$ and $\epsilon^-_\bz=\epsilon^+_z$ ensures that the anti-chiral part of the charge is also finite.
This yields the following symplectic structure
\be
\begin{aligned}\label{Omegasubleadingchi}
\Omega(\delta\chi,\delta'\chi)={i} \int_{\scri^+} du d^2z &\Big\{\sqrt{\gamma}\left[\delta\chi^{\dagger\, (2)}_r\left(\frac{1}{2}\sb_r-\sb_u\right)\delta'\chi^{(0)}_u+\delta\chi^{\dagger\, (1)}_u\left(\frac{1}{2}\sb_r-\sb_u\right)\delta'\chi^{(1)}_r\right]\\
&\qquad -\delta\chi^{\dagger\, (1)}_\bz \left(\frac{1}{2}\sb_r-\sb_u\right)\delta'\chi^{(-1)}_z+\delta\chi_z^{\dagger\, (0)}\sb_u\delta\chi^{(0)}_\bz \Big\}+ h.c.\,.
\end{aligned}
\ee
In section~\ref{sec:subgravitino} we are interested in the case where $\delta'\chi$ is replaced by the subleading conformally soft gravitino primary.
 \\

\bibliographystyle{utphys}
\bibliography{PPP}

\end{document}